\documentclass[11pt,a4paper]{article}
\usepackage{jheppub}
\usepackage{amsmath,epsfig}
\usepackage{amssymb,amsfonts}
\usepackage{latexsym}
\usepackage{epsfig}
\usepackage{pdfsync}
\usepackage{float}
\newbox\pippobox
\def\be{\begin{equation}}
\def\ee{\end{equation}}
\def\bea{\begin{eqnarray}}
\def\eea{\end{eqnarray}}

\def\6{\partial}

\def\a{\alpha}

\def\C0{{\bf C_0}}
\def\Y0{{\bf Y_0}}
\def\G0{{\bf G_0}}

\def\sq
\def\a{\alpha}

\title{Chiral and Deconfining Phase Transitions from Holographic QCD Study}
\author[a,b]{Zhen Fang,}
\author[c,d,a]{Song He,}
\author[a]{Danning Li}

\affiliation[a]{State Key Laboratory of Theoretical Physics,
Institute of Theoretical Physics, Chinese Academy of Science,
Beijing 100190, P. R. China }
\affiliation[b]{University of Chinese Academy of Sciences,
Beijing, 100190, P.R. China}
\affiliation[c]{Max Planck Institute for Gravitational Physics (Albert Einstein Institute)
Am M\"{u}hlenberg 1, 14476 Golm, Germany}
\affiliation[d]{Yukawa Institute for Theoretical Physics, Kyoto University, Kitashirakawa Oiwakecho,\\
Sakyo-ku, Kyoto 606-8502, Japan}
\emailAdd{fangzhen@itp.ac.cn}\emailAdd{hesong17@gmail.com}\emailAdd{lidn@itp.ac.cn}

\date{\today}

\abstract{ A first attempt to accommodate the chiral and deconfining phase transitions of QCD in the bottom-up holographic framework is given. We constrain the relation between dilaton field $\phi$ and metric warp factor $A_e$ and get several reasonable models in the Einstein-Dilaton system. Using the potential reconstruction approach, we solve the corresponding gravity background. Then we fit the background-related parameters by comparing the equation of state with the two-flavor lattice QCD results. After that we study the temperature dependent behavior of Polyakov loop and chiral condensate under those background solutions. We find that the results are in good agreement with the two-flavor lattice results. All the studies about the equation of state, the Polyakov loop and the chiral condensate signal crossover behavior of the phase transitions, which is consistent with the current understanding on the QCD phase transitions with physical quark mass. Furthermore, the extracted transition temperatures are comparable with the two-flavor lattice QCD results. }


\begin{document}

\maketitle


\section{Introduction}

Spontaneous chiral symmetry breaking and color confinement are the two most important properties of the vacuum of Quantum Chromodynamics (QCD), which is widely accepted as the fundamental theory of the strong interaction. At sufficient high temperature and/or density, it is believed that phase transition might happen in the system, including the restoration of chiral symmetry and the release of color degrees of freedom. At present, to understand the phase structure of these two phase transitions is attracting more and more attention in both non-perturbative QCD study and cosmology \cite{nature-PTD}.

Generally, the properties of the two phase transitions would depend sensitively on the intrinsic quantities of the system. For example, the chiral phase transition is well defined as a true phase transition only in the chiral limit, i.e., zero quark mass limit, while the deconfining phase transition should be in a totally opposite limit, i.e., the infinite quark mass limit. This is because only in these two limits the chiral symmetry and $Z_3$ center symmetry, the breaking and restoration of which are related to the phase transitions, become the exact symmetries of QCD. In these limits, chiral condensate $\langle\bar{\psi}\psi\rangle$ and Polyakov loop $\langle L \rangle$ could be well defined as order parameters for chiral and deconfining phase transition respectively. In the physical quark mass region, there are no exact symmetries and the phase transition might turn to a rapid but continuous crossover transition.

\begin{figure}[!h]
\begin{center}
\epsfxsize=7.5 cm \epsfysize=7.5 cm \epsfbox{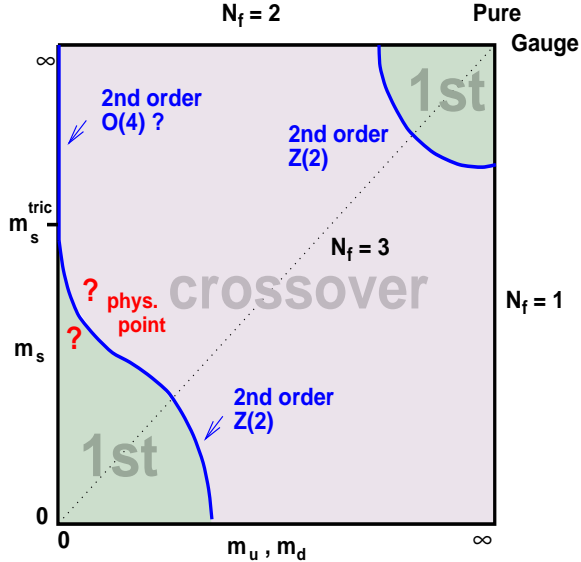} \
\end{center}
\caption{The expected phase diagram in the current quark mass space with degenerate u,d quark masses (Taken from \cite{qcd-phase-diagram}).} \label{columbia-plot}
\end{figure}

Based on theoretical consideration and lattice QCD simulation \cite{qcd-phase-diagram,deForcrand:2006pv,Kanaya:2010qd}, a possible $2+1$ flavor phase diagram in the current quark mass plane is summarized in the sketch (sometimes called "Colombia Plot") shown in Fig.\ref{columbia-plot} \cite{qcd-phase-diagram}. In this plot, there are two regions with first-order phase transition, i.e., near the chiral limit region ($m_u=m_d=m_s\simeq0$) and near the infinite quark mass region ($m_u=m_d=m_s\simeq\infty$). In the intermediate region it is expected to be a crossover transition. There are two second-order lines as the boundaries between the first-order regions and the crossover region. It is noted that in the region of two light flavors where $m_u=m_d\simeq O(\text{MeV})$ and $m_s=\infty$, even very small quark mass would drive the second-order transition in the chiral limit to a crossover transition with finite quark mass (in analogy to the $O(4) ~ \sigma$ model \cite{Pisarski:1983ms} noting that $SU(2)_L\times SU(2)_R\simeq O(4)$). In this paper, as a preliminary try, we will focus on the behavior of phase transitions in this area.

In addition to the quark mass and flavor dependent behavior of QCD phase transition, one of the most important things for both experimental investigations of quark gluon plasma and theoretical studies of thermal QCD is to estimate the transition temperature $T_c$. As the improvement of lattice computation in recent years, most of the results about the transition temperatures converge towards $T_c\simeq145\sim165~\rm{MeV}$ in $2+1$ flavor QCD \cite{Kanaya:2010qd}. In the two-flavor case, the transition temperature is around $170~\rm{MeV}$ when extrapolated to the chiral limit, while in the three-flavor case it is about $155~\rm{MeV}$ \cite{qcd-phase-diagram}. The transition temperature would increase with increasing quark mass. Another interesting topic is about the relation of chiral and deconfining transitions. In this aspect, discrepancies exist for different theoretical studies. Besides lattice simulations (see references in \cite{Kanaya:2010qd}), there are also other efforts such as using effective models or by functional methods \cite{Fischer:2009gk,Fischer:2011mz,Xu:2011pz,Andersen:2013swa}. In this work, we will try to use the holographic approach to study the thermal phase transition of QCD and try to provide more understanding on it.

In recent decades, the discovery of the anti-de Sitter/conformal field theory (AdS/CFT) correspondence and the conjecture of the gravity/gauge duality \cite{Maldacena:1997re,Gubser:1998bc,Witten:1998qj} shed new light on the strong coupling problem of gauge theory. Based on this idea, people have extended it to study gauge theory like QCD, in which conformal symmetry is broken dynamically at low energy. By breaking the conformal symmetry in different ways, many efforts have been made towards more realistic holographic description of the low energy phenomena of QCD, such as in hadron physics \cite{Erlich:2005qh,TB:05,DaRold2005,D3-D7,D4-D6,SS-1,SS-2,Csaki:2006ji,Dp-Dq,Gherghetta-Kapusta-Kelley,Gherghetta-Kapusta-Kelley-2,YLWu,YLWu-1,Cui:2013xva,Li:2012ay,Li:2013oda} and hot/dense QCD \cite{Shuryak:2004cy,Tannenbaum:2006ch,Policastro:2001yc,Cai:2009zv,Cai:2008ph,Sin:2004yx,Shuryak:2005ia,Nastase:2005rp,Nakamura:2006ih,Sin:2006pv,
Janik:2005zt,Herzog:2006gh,Gubser-drag,Zhang:2012jd,Wu:2014gla,Li:2014hja,Li:2014dsa}, both in top-down approaches and in bottom-up approaches (see \cite{Aharony:1999ti,Erdmenger:2007cm,Brodsky:2014yha,Kim:2012ey,Adams:2012th} for reviews).

For QCD phase transitions, most of the bottom-up studies \cite{Herzog:2006ra,BallonBayona:2007vp,Kim:2007em,Kim:2009wt,Cai:2007zw,Cai:2012eh,Andreev-T3,Colangelo:2010pe,Gubser:2008ny,Gubser:2008sz,Gubser:2008yx,Gursoy:2008bu,Gursoy,Gursoy-3,Gursoy:2008za,Finazzo:2014zga,Yaresko:2015ysa,Yaresko:2013tia,Li:2011hp,Cai:2012xh,He:2013qq,Yang:2014bqa,Yang:2015aia,Zuo:2014iza,Zuo:2014vga,Afonin:2014jha,Cui:2014oba} focus only on deconfining phase transition. To add chiral aspects, the soft-wall model \cite{Karch:2006pv} provides a starting point, since the model itself and its extended ones \cite{Gherghetta-Kapusta-Kelley,Gherghetta-Kapusta-Kelley-2,YLWu,YLWu-1,Cui:2013xva,Li:2012ay,Li:2013oda} have been shown to characterize the hadron spectra and other quantities quite well. Furthermore, in these models the chiral condensate, which is the order parameter of chiral phase transition, was introduced to realize the spontaneous chiral symmetry breaking of QCD vacuum. However, unlike the Nambu-Jona-Lasinio (NJL) model \cite{Nambu:1961tp,Nambu:1961fr}, the value of the chiral condensate was often taken as a free parameter to fit the hadron spectra at zero temperature instead of being self-consistently solved from the model itself. Noting that the IR boundary condition may require the dependence of chiral condensate on quark mass, the authors of \cite{Colangelo:2008us} extended the soft-wall model to the finite temperature case and solved the temperature dependent chiral condensate self-consistently. In the previous work \cite{Chelabi:2015cwn,chiral-long}, we extended their method and tried to get more constraints on the model from chiral aspects of QCD phase transition. There we showed how to get the correct mass dependent behavior of chiral phase transition as shown in Fig.\ref{columbia-plot}. Furthermore, we also see that under the AdS-Schwarzchild (AdS-SW) black hole background new constraints on the dilaton field come up: it should be negative at certain UV scale in addition to the IR constraints from meson spectra. However, since in that work we only focused on the chiral phase transition and used the AdS-SW black hole solution as the background geometry, we did not consider the deconfining phase transition there. In this work, we will try to grasp the main requirement in describing chiral and deconfining phase transition simultaneously, and a holographic QCD (hQCD) model will be built up to characterize the behavior of the two aspects of phase transitions.

The paper is organized as follows. In Sec.\ref{section-graviton-dilaton}, we describe the Einstein-dilaton system, which has been used to characterize the deconfining phase transition in previous studies. We try to use the potential reconstruction method to construct several models. Then we study thermodynamics of these models and compare the results with two-flavor lattice QCD simulations in Sec.\ref{eos}. After fixing the parameters by the description of equation of state, we study the temperature dependent behavior of Polyakov loop and chiral condensate, and also compare them with the results from lattice in Sec.\ref{phasetransition}. In Sec.\ref{section-summary}, we give a short discussion and conclusion.

\section{Gravity setup}\label{ED}
\label{section-graviton-dilaton}

As mentioned above, in bottom-up holographic studies, the deconfining phase transition has already been widely investigated in the Einstein-dilaton system \cite{Gubser:2008ny,Gubser:2008sz,Gubser:2008yx,Gursoy:2008bu,Gursoy,Gursoy-3,Gursoy:2008za,Finazzo:2014zga,Yaresko:2015ysa,Yaresko:2013tia,Li:2011hp,Cai:2012xh,He:2013qq,Yang:2014bqa,Yang:2015aia,Zuo:2014iza,Zuo:2014vga}. In addition, in \cite{Chelabi:2015cwn,chiral-long} we showed the possibility to characterize chiral symmetry breaking and its restoration in soft-wall model. In light of these researches, we expect that by combining these two systems it should be possible to describe the two most important aspects of QCD phase transition simultaneously.  Therefore, we consider the following action in string frame
\begin{eqnarray}
S &=& S_b+S_m, \label{whole action}\\
S_b &=&\frac{1}{16\pi G_5}\int d^5x\sqrt{-g^S}e^{-2 \phi }\left[R^S+4\partial_\mu \phi \partial ^\mu \phi - V_S(\phi)\right], \label{background action string frame}\\
S_m &=&-\int d^5x
 \sqrt{-g^S}e^{-\phi}\rm{Tr}\left[\nabla_{\mu}X^{\dagger} \nabla^{\mu}X+V_X(|X|)\right]\label{matter action},
\end{eqnarray}
where $S_b$ is the background sector and $S_m$ the matter sector. The index S in the integrand denotes the string frame and $G_5$ is the 5D Newton constant. There are two scalar fields, i.e., the dilaton field $\phi$ and the bulk scalar field $X$ which is dual to the $\bar{q}q$ condensate of QCD. $V_S$ and $V_X$ represent the dilaton potential and the bulk scalar potential respectively. Here the leading term of $V_X$ is the mass term $m_5^2 XX^{\dagger}$ and the bulk scalar mass $m_5^2=-3$ can be determined from the AdS/CFT prescription $m_5^2L^2=(\Delta-p)(\Delta+p-4)$ by taking $\Delta=3, p=0$ \cite{Witten:1998qj}.

In \cite{Li:2011hp,Cai:2012xh}, we showed that the Einstein-dilaton system can describe pure gluon thermodynamics quite well. After adding the flavor sector $S_m$, we found that the above system can describe the meson spectra which are consistent with experimental data \cite{Li:2012ay,Li:2013oda}. However, extending this model to the finite temperature case and trying to solve this gravity-two-scalar coupled system is quite a complicated work. Actually, in \cite{Gubser:2008ny,Gubser:2008yx,Finazzo:2014zga}, the authors also tried to study thermodynamics of QCD with flavors in the Einstein-dilaton system. Following this logic and as a preliminary try, we will solve the Einstein-dilaton action $S_b$ as the geometrical background and take the matter action $S_m$ as a probe, which should be considered as an approximation of the full system. As pointed out above, we consider chiral and deconfining phase transitions in the two-flavor case ($N_f=2$), so the background geometry will be constrained by $N_f=2$ thermodynamics before studying the temperature dependent behavior of chiral condensate and Polyakov loop. In this section, we will first outline the necessary framework of the Einstein-dilaton system, and then try to constrain the background and give several models for study.

\subsection{Equation of motion for background geometry}

Given the background action $S_b$ as shown in Eq.(\ref{background action string frame}), if one knows the dilaton potential $V_S(\phi)$, then the whole system could be solved numerically. This is the usual approach to deal with this system, as can be seen in \cite{Gubser:2008ny,Gubser:2008sz,Gubser:2008yx,Gursoy:2008bu,Gursoy,Gursoy-3,Gursoy:2008za,Finazzo:2014zga,Yaresko:2015ysa,Yaresko:2013tia}, where by tuning the dilaton potential carefully the authors can describe the QCD equation of state quite well. While in \cite{Li:2014hja,Li:2014dsa,Li:2011hp,Cai:2012xh,He:2013qq}, we used an approximate approach, which is usually called potential reconstruction approach, to construct the geometrical background (see also \cite{He:2010ye,He:2011hw,Yang:2014bqa,Yang:2015aia}). In this approach, once fixing the dilaton profile \cite{Li:2014hja,Li:2014dsa}, the metric warp factor \cite{Li:2011hp,Cai:2012xh,He:2013qq,Yang:2014bqa,Yang:2015aia} or the relations between them, the dilaton potential could be solved from the equations of motion, which nevertheless entails a temperature dependence of the potential. However, in the region concerned, the temperature dependence of dilaton potential is very weak, so it can be seen as an approximation of the potential fixing approach (see also the discussion in \cite{Kajantie:2011nx}). Furthermore, it turns out to be easier to generate the background solution. Hence, in this work, we will use this approach and try to get the necessary ingredients in describing QCD phase transitions holographically. Here, we first review how to use the potential reconstruction approach to obtain solutions in the 5D Einstein-dilaton system given in Eqs.(\ref{whole action})-(\ref{matter action}).

For convenience, we give the relevant formulas in Einstein frame. By conformal transformation of the metric
\begin{equation}\label{metric trans}
g_{\mu \nu}^S=e^{\frac{4\phi}{3}} g_{\mu \nu}^E,
\end{equation}
the action in Einstein frame is derived as
\begin{equation}\label{background action Einstein frame}
S_b=\frac{1}{16\pi G_5}\int d^5x\sqrt{-g^E}\left[R^E-\frac{4}{3}\partial_\mu \phi \partial ^\mu \phi - V_E(\phi)\right],
\end{equation}
with
\begin{equation}\label{dilaton potential}
V_E(\phi)=e^{\frac{4\phi}{3}} V_S(\phi).
\end{equation}

For finite temperature solution, the metric ansatz in string frame and Einstein frame will be taken as follows
\begin{eqnarray}
ds_S^2 &=& \frac{L^2 e^{2 A_s}}{z^2} \left(-f(z)dt^2 + dx^i dx^i + \frac{dz^2}{f(z)}\right)   \label{stringmetric},\\
ds_E^2 &=& \frac{L^2 e^{2 A_e}}{z^2} \left(-f(z)dt^2 + dx^i dx^i + \frac{dz^2}{f(z)}\right)   \label{Einsteinmetric},
\end{eqnarray}
where $L$ is the radius of $\rm{AdS}_5$, and the relation of the metric warp factor in different frames is $A_s=A_e+2\phi/3$.

From the action (\ref{background action Einstein frame}), the general Einstein equation can be derived as
\begin{equation}\label{Einstein equation}
E_{\mu \nu}+\frac{1}{2}g^E_{\mu \nu}\left(\frac{4}{3}\partial_\mu \phi \partial^\mu \phi +V_E(\phi)\right)-\frac{4}{3}\partial_\mu \phi \partial_\nu \phi=0,
\end{equation}
from which we obtain the non-zero components:
\begin{eqnarray}
A_e''(z)+A_e'(z)\left(\frac{f'(z)}{2 f(z)}-\frac{2}{z}+A_e'(z)\right)-\frac{f'(z)}{2 z f(z)}+\frac{2}{z^2}+\frac{2}{9} \phi '(z)^2 & &\nonumber \\
+\frac{L^2 e^{2 A_e(z)} V_E(\phi(z))}{6 z^2 f(z)}&=&0, \label{tt}
\end{eqnarray}
\begin{eqnarray}
-\frac{9 f'(z) A_e'(z)}{4 f(z)}-9 A_e'(z){}^2+\frac{18 A_e'(z)}{z}+\frac{9 f'(z)}{4 z f(z)}-\frac{9}{z^2}+\phi '(z)^2 & & \nonumber \\
-\frac{3 L^2 e^{2 A_e(z)} V_E(\phi(z))}{4 z^2 f(z)} &=&0,  \label{zz}
\end{eqnarray}
\begin{eqnarray}
f''(z) + f(z)\left(6 A_e'(z){}^2-\frac{12 A_e'(z)}{z}+6 A_e''(z)+\frac{12}{z^2}+\frac{4}{3} \phi '(z)^2\right) & &\nonumber \\
+ f'(z)\left(6 A_e'(z)-\frac{6}{z}\right)+\frac{L^2 e^{2 A_e(z)} V_E(\phi (z))}{z^2} &=&0. \label{xx}
\end{eqnarray}
Note that we only need two of the above three equations in which $V_E(\phi)$ is considered as a derived quantity in the potential reconstruction approach. The other one can be used as a consistent check for solutions of the equations of motion. For simplicity, we recombine Eqs.(\ref{tt})-(\ref{xx}) and obtain the following two simplified equations:
\begin{eqnarray}
f''(z)+3 f'(z)\left(A_e'(z)-\frac{1}{z}\right) &=&0, \label{ff}\\
A_e''(z)+A_e'(z)\left(\frac{2}{z}-A_e'(z)\right)+\frac{4}{9} \phi'(z)^2 &=&0. \label{AE}
\end{eqnarray}
Also from the action (\ref{background action Einstein frame}), we obtain the dilaton field equation:
\begin{equation} \label{dilaton}
\phi''(z)+\phi'(z) \left(3 A_e'(z)+\frac{f'(z)}{f(z)}-\frac{3}{z}\right)-\frac{3 L^2 e^{2 A_e(z)} \partial_\phi V_E(\phi (z))}{8 z^2 f(z)}=0.
\end{equation}

Note that only $A_e$ and $\phi$ appear in Eq.(\ref{AE}). If one of these two quantities or the relation between them were given, they could be solved from Eq.(\ref{AE}), then $f$ and $V_E(\phi)$ would be obtained by solving Eq.(\ref{ff}) and Eq.(\ref{dilaton}). Usually only the integral constant in $f$ would appear in the final expression of $V_E(\phi)$. As the integral constant is related to the black hole temperature, this indicates that $V_E(\phi)$ would depend on temperature. However, as we can see later, the temperature dependence of $V_E(\phi)$ is very weak, which makes it possible to consider this approach as an approximation of fixing dilaton potential. Taking specific profile of $\phi$ and $A_e$, we studied thermodynamics of the pure gluon system in \cite{Li:2011hp,Cai:2012xh,Li:2014dsa,Li:2014hja} and found that the results are in good agreement with the quenched lattice QCD results. In this work, we aim at characterizing thermodynamics and phase transitions in $N_f=2$ QCD with finite quark mass, which shows a crossover transition instead of a first-order one such as appears in the pure gluon system. Thus, we will attempt to acquire new constraints on $A_e,\phi$ and construct the gravity background for characterizing two-flavor QCD thermodynamics.

\subsection{UV and IR constraints on background geometry and dilaton} \label{phi-A}

To tackle the gravity background, $\phi$ and $A_e$ (or equivalently $A_s$) should be specified as the input of the background Eqs.(\ref{ff})-(\ref{dilaton}). Starting from a specific relation, an asymptotic AdS black hole solution will be obtained. In this section, we will first find the UV and IR constraints of the corresponding quantities, and then try to build reasonable gravity background models. In our convention, the UV and IR region are corresponding to $z\sim0$ and large $z$ respectively.

Firstly, as noted in \cite{Li:2012ay,Li:2013oda,Li:2011hp,Cai:2012xh}, the Einstein-dilaton system should be closely related to the gluon dynamics, which means that the dimension of the dilaton field should be $\Delta=2$ or $\Delta=4$, which is equivalent to requiring the leading UV asymptotic behavior of $\phi$ to be $z^2$ or $z^4$ forms according to the holographic dictionary.

Secondly, as shown in \cite{Li:2011hp,Cai:2012xh}, the UV asymptotic AdS region is related to the high temperature behavior of the thermodynamic system. The asymptotic AdS property guarantees the system to approach conformal gas at very high temperature. However, in order to have a correct description of the low temperature behavior, the IR asymptotic behavior of the background fields should be carefully tuned. From the previous studies \cite{Li:2012ay,Li:2013oda,Li:2011hp,Cai:2012xh}, we see that if the IR behavior of $A_s$ or $\phi$ is of the quadratic form $z^2$, then the black hole solutions would have a minimal temperature and the system shows a first-order phase transition. However, if the flavor sector and quark mass are taken into consideration, the QCD phase transition will turn into a crossover one without a real phase transition. Therefore, we need a gravity background which can link the high temperature phase with the low temperature phase. In this case, we tune the IR behavior of the fields carefully and find that when they approach a constant at IR the temperature corresponding to the AdS black hole solution can goes down to zero continuously, as will be shown in Fig.\ref{Tzh}.\footnote{Here we note that this requirement is in contradiction with constraints from meson spectrum, where an IR quadratic dilaton is needed to produce the Regge behavior of meson spectra. However, in this manuscript, we just want to grasp the most important ingredients in describing QCD phase transitions and not to lay emphasis on the mass spectrum. Furthermore, we note that in \cite{Batell:2008zm}, by adding an extra scalar field, it is possible to get pure AdS together with IR quadratic dilaton field which is consistent with the spectrum calculation. One might combine these two aspects in one model through this way. We will leave the more careful study to the future.}

Thirdly, the main motivation of this paper is to study deconfining phase transition together with chiral phase transition. In our previous study \cite{Chelabi:2015cwn,chiral-long}, we found that new constraints on dilaton field could be obtained from the chiral aspects of QCD phase transition. As here the gravity background is no longer pure AdS, One needs another similar constraint on $5A_s-\phi$, noting that $e^{5A_s-\phi}$ comes from the sector $\sqrt{-g^S}e^{-\phi}=\frac{1}{z^5}e^{5A_s-\phi}$ of the matter action (\ref{matter action}) which will be considered in Sec.\ref{chiral part}. In \cite{Chelabi:2015cwn,chiral-long} we showed that a negative part of dilaton at certain scale not far from UV is necessary to obtain the correct chiral phase transition behavior in the pure AdS background. Accordingly, here we only require $5A_s-\phi$ to be positive for simplicity.

As a short summary, we need a UV leading $z^2$ or $z^4$ configuration of $5A_s-\phi$ or $\phi$ itself, and we will simply set the IR behavior of them to approach a positive constant for a possible crossover transition. The simplest choice to interpolate the UV and IR behavior required above is of $tanh$ form, and the next order term will be retained to fit the correct intermediate behavior of thermal transition. Hence, we give the following four possible ansatz, and set them to be model A1 (A2) and model B1 (B2),
\begin{eqnarray}
5 A_s(z)-\phi(z) &=  \beta\,\mathrm{tanh}(\mu^2 z^2+\nu^4 z^4)    &\qquad\text(\mathrm{model\,A1}) \label{ph-A relation1},\\
\phi(z) &=  \beta\,\mathrm{tanh}(\mu^2 z^2+\nu^4 z^4)             &\qquad\text(\mathrm{model\,A2}) \label{ph-A relation2},\\
5 A_s(z)-\phi(z) &=  \beta\,\mathrm{tanh}(\mu^4 z^4+\nu^6 z^6)    &\qquad\text(\mathrm{model\,B1}) \label{ph-A relation3},\\
\phi(z) &=  \beta\,\mathrm{tanh}(\mu^4 z^4+\nu^6 z^6)             &\qquad\text(\mathrm{model\,B2}) \label{ph-A relation4}.
\end{eqnarray}
Here $\beta,\mu,\nu$ are model parameters and will be fixed later by comparing the results of equation of state with those from lattice simulations.

It is easy to obtain the UV asymptotic behavior of $\phi(z)$ under the above settings as follows
\begin{eqnarray}
\label{asybehavior1} \phi(z\rightarrow 0) & \sim \frac{3\beta \mu^2 z^2}{7} + \cdots      &\qquad\text(\mathrm{model\,A1}),\\
\label{asybehavior2} \phi(z\rightarrow 0) & \sim \beta \mu^2 z^2 + \cdots                 &\qquad\text(\mathrm{model\,A2}),\\
\label{asybehavior3} \phi(z\rightarrow 0) & \sim \frac{3\beta \mu^4 z^4}{7} + \cdots      &\qquad\text(\mathrm{model\,B1}),\\
\label{asybehavior4} \phi(z\rightarrow 0) & \sim \beta \mu^4 z^4 + \cdots                 &\qquad\text(\mathrm{model\,B2}).
\end{eqnarray}
From these results, we see that the conformal dimension of the dilaton field is $\Delta=2$ in A1 (A2) model and $\Delta=4$ in B1 (B2) model and they all satisfy the BF bound. The dilaton field with $\Delta=4$ can be dual to the gauge invariant dimension-4 gluon condensate, while the dilaton with $\Delta=2$ does not correspond to any local, gauge invariant operator in QCD. Although there have been many discussions in recent years of the possible relevance of a dimension-two condensate in the form of a gluon mass term \cite{Gubarev:2000eu,Gubarev:2000nz,Kondo:2001nq}, it is still not clear whether we can associate $\phi$ with a dimension-2 operator, because the AdS/CFT correspondence requires that the bulk fields should be dual to gauge-invariant local operators. Despite of all these issues, we will just go on to find the physical implications on the thermal QCD phase transition from the holographic studies.

\section{Equation of state for the hQCD model}
\label{eos}

After giving the main framework of the hQCD model, in this section, we will investigate the equation of state in these models given in Eqs.(\ref{ph-A relation1})-(\ref{ph-A relation4}). We fix the parameters $\beta,\mu,\nu$ and $G_5$ in the models by comparing the calculated results with the $N_f=2$ lattice ones from \cite{Burger:2014xga}, and list them in Table.\ref{parameter value} for later use.
\begin{table}[!h]
\begin{center}
\begin{tabular}{cccccccccc}
\hline\hline
                 & $\beta$  &$\mu(\rm{GeV})$    &    $\nu(\rm{GeV})$ &    $G_5$        \\
\hline
       Model A1  & 2.5      & 0.5           &      0.46      &     1.06         \\

       Model A2  & 1.8      & 0.4           &       0.42     &     1.06          \\

       Model B1  & 1.8      & 0.59          &        0.52    &     1.1             \\

       Model B2  & 1.3      & 0.55          &        0.38    &     1.1             \\
\hline\hline
\end{tabular}
\caption{Value of parameters in model A1 (A2) and model B1 (B2).}
\label{parameter value}
\end{center}
\end{table}

\subsection{Black hole solutions and associated thermodynamics}
\label{blackhole}

Before going to the details, we will list general formulas of some thermodynamic quantities. Here we are interested in a series of solutions whose UV behavior is asymptotic $\rm{AdS}_5$. We also impose the requirements: $f(0)=1$, and $\phi(z), f(z), A_s(z)$ are regular from $z=0$ to $z=z_h$. Here $z_h$ is the black hole horizon with $f(z_h)=0$.

The Hawking temperature of the black hole solution is defined by
\begin{equation}\label{temp}
T=\frac{|f'(z_h)|}{4\pi}.
\end{equation}
The background Eq.(\ref{ff}) can be simplified by defining $f(z)$ as
\begin{equation} \label{ffhh}
f(z)=1-\frac{h(z)}{h(z_h)}
\end{equation}
with $h(0)=0$. Then the temperature formula (\ref{temp}) becomes
\begin{eqnarray} \label{T2}
T=\frac{h^{'}(z_h)}{4\pi h(z_h)}.
\end{eqnarray}
Substituting Eq.(\ref{ffhh}) in Eq.(\ref{ff}), we obtain a simplified equation of $h(z)$:
\begin{equation} \label{hh}
h'(z)-e^{-3A_e(z)}z^3 = 0.
\end{equation}

\begin{figure}[!h]
\begin{center}
\epsfxsize=7 cm \epsfysize=4.5 cm \epsfbox{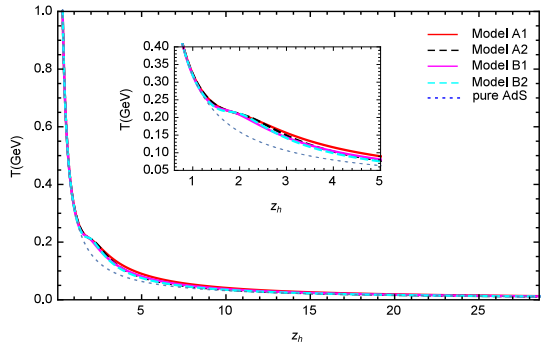} \vskip -0.5 cm
\end{center}
\caption{The temperature as a function of horizon $z_h$ in model A1 (A2) and model B1 (B2).}
\label{Tzh}
\end{figure}

Now we are ready to solve the background system completely. In Fig.\ref{Tzh}, we present the temperature as a function of the horizon $z_h$ in model A and B. We see that the temperature decrease monotonously with $z_h$ from high ones to zero. The fact that the behavior of $T(z_h)$ is almost same in all the four models quantitatively means that the black hole temperature is not sensitive to the relations of $A_s$ and $\phi$ we have chosen. That is critical for a consistent realization of the crossover transition and also important as an indication that the model is robust enough in characterizing the transition features of thermal QCD. Compared with the pure AdS-SW black hole case, there is a small $z_h$ region in which the temperature decrease slightly slowly in these models, as can be seen from the insert of Fig.\ref{Tzh}.

\begin{figure}[!h]
\begin{center}
\epsfxsize=7 cm \epsfysize=4.5 cm \epsfbox{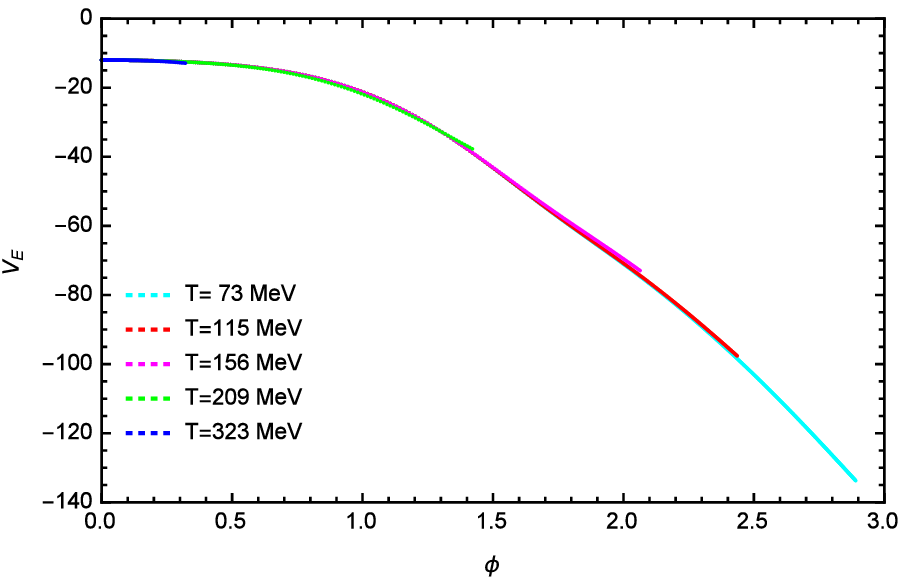}
\hspace*{0.6cm} \epsfxsize=7 cm \epsfysize=4.5 cm
\epsfbox{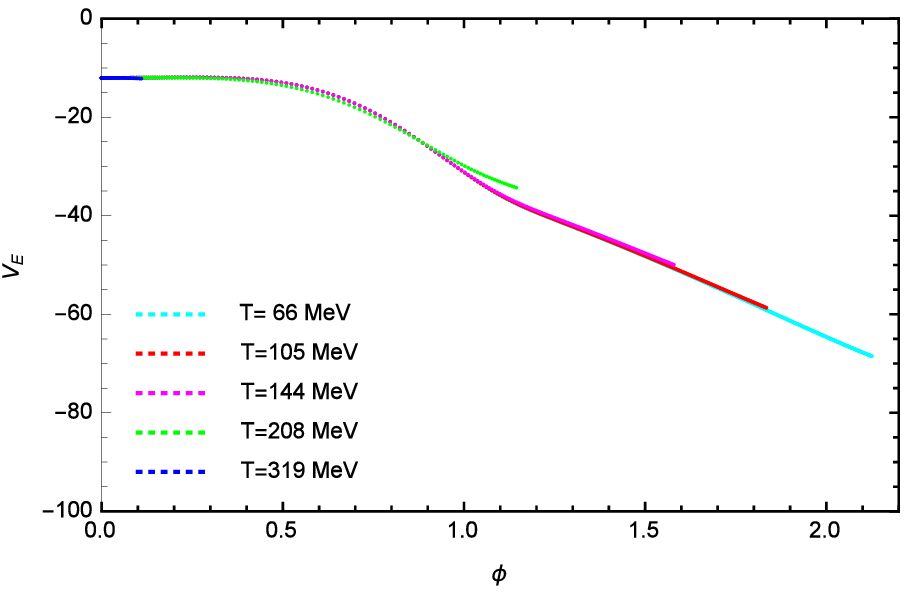} \vskip -0.05cm \hskip 0.7 cm
\textbf{(a)} \hskip 7.2 cm \textbf{(b)} \\
\end{center}
\caption{The dilaton potential in model A1 (a) and model B1 (b).}
 \label{Vph-AB}
\end{figure}

The dilaton potential can be derived from Eqs.(\ref{tt})-(\ref{xx}), and the results in model A1 and B1 are shown in Fig.\ref{Vph-AB} with several different temperatures. From Fig.\ref{Vph-AB} one can see that the dilaton potential will approach to the negative cosmological constant when $\phi$ goes to zero, which is consistent with the UV asymptotic AdS boundary. In the IR region, the potential will be dependent on the temperature due to the potential construction. However, since the relevant physical region is from the boundary $z=0$ to the horizon $z=z_h$, we only plot the section from $\phi=0$ to $\phi=\phi(z_h)$. One can see that in this region the dilaton potential almost does not change with temperature. In this sense, it might be possible to build up a model with fixing dilaton potential which could produce similar results as that in this work.

\subsection{Entropy density and sound speed}

Based on the classical Bekenstein-Hawking formula and using the metric ansatz (\ref{Einsteinmetric}) in Einstein frame, the formula of the black-hole entropy density is derived as
\begin{equation} \label{entropy}
 s=\left.\frac{A_{area}}{4 G_5 V_3}\right|_{z_h}=\left.\frac{L^3}{4G_5}\frac{e^{3A_e}}{z^3}\right|_{z_h}
\end{equation}
with $A_{area}$ the area of horizon, $G_5$ the 5D Newton constant and $V_3$ the space volume. The sound speed is defined as
\begin{equation} \label{soundspeed}
c_s^2=\frac{d\,\mathrm{ln}T}{d\,\mathrm{ln}s}.
\end{equation}

\begin{figure}[!h]
\begin{center}
\epsfxsize=7 cm \epsfysize=4.5 cm \epsfbox{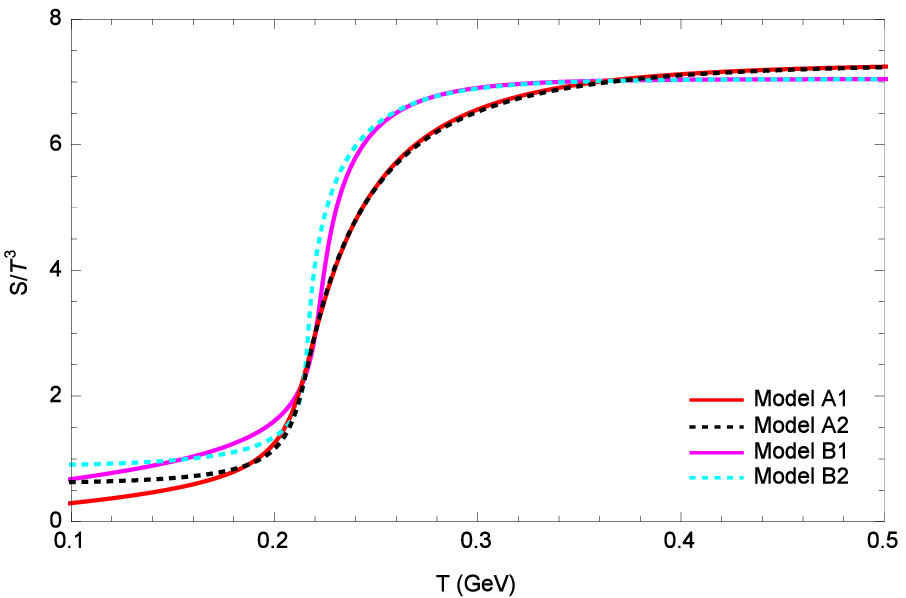}
\hspace*{0.6cm} \epsfxsize=7 cm \epsfysize=4.5 cm
\epsfbox{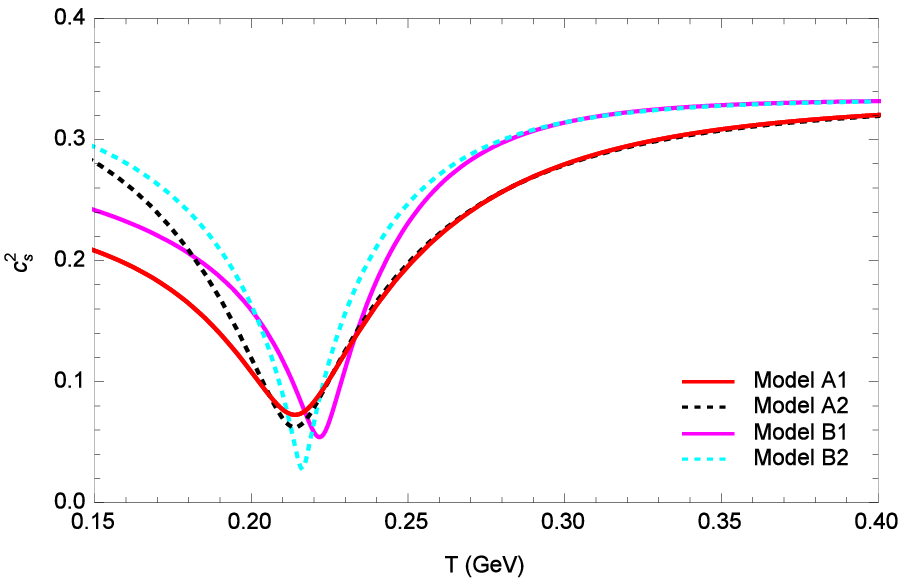} \vskip -0.05cm \hskip 0.7 cm
\textbf{(a)} \hskip 7.2 cm \textbf{(b)} \\
\end{center}
\caption{Temperature dependence of the scaled entropy density $s/T^3$ (a) and the square of sound speed $c_s^2$ (b) in model A1 (A2) and model B1 (B2).}
 \label{s-cs2}
\end{figure}

Given the formula of temperature $(\ref{temp})$ and entropy density $(\ref{entropy})$, the speed of sound can be obtained. It has been well known that for conformal system, $c_s^2 = 1/3$, while for non-conformal system, $c_s^2$ will deviate from $1/3$. From Eq.(\ref{soundspeed}), one can see that the speed of sound is independent of the normalization of the 5D Newton constant $G_5$ and the space volume $V_3$. Fig.\ref{s-cs2} presents the scaled entropy density $s/T^3$ and the sound speed $c_s^2$ as a function of $T$ in model A1 (A2) and model B1 (B2).

One can see from Fig.\ref{s-cs2} (a) that the crossover transition is obvious, and the scaled entropy density increases with temperature. There is a small temperature region where $\frac{s}{T^3}$ increases rapidly, which indicates a sudden release of degrees of freedom. At high temperature, $\frac{s}{T^3}$ approaches to a finite value $\frac{\pi^3}{4G_5}$ which can be obtained from asymptotic analysis. From thermal QCD one knows that the scaled entropy density at high temperature would approach the Stefan-Boltzmann limit which is related to the degrees of freedom of the system. Thus, in this sense, if we tune $G_5$ and fix the high temperature value of $\frac{s}{T^3}$ comparable with the $N_f=2$ lattice QCD result, the flavor effects will partly incorporated. In Fig.\ref{s-cs2} (b), we plot the square of sound speed varying with temperature. All these four models show a minimal value within the temperature region $210\sim 230~\rm{MeV}$. This also indicates a crossover behavior which is consistent with the entropy density. At high temperature, the square of sound speed goes to the conformal value $1/3$. It should be noted that our high temperature analysis of the thermal quantities may be irrelevant from the point of view of holography because at very high temperatures the property of asymptotic freedom will be attained, which makes the low energy supergravity approximation of the full string theory used in the AdS/CFT (weak/strong) duality invalid. However, from the results we obtained, one finds that the high temperature behavior is consistent with the lattice or other thermal QCD results, at least in a temperature region (up to $0.4~$GeV) accessible to the lattice simulations. So we assume that when the temperature is not far below or far above the thermal transition region, the hQCD model can give an approximately good description for the thermal QCD transition.

\subsection{Pressure density, energy density and trace anomaly}

The pressure density $p$ is given by the formula
\begin{equation} \label{p-T}
\frac{dp}{dT}= s.
\end{equation}
Numerically, we transform Eq.(\ref{p-T}) into $p'(z_h)=s(z_h) T'(z_h)$ which is solved by giving the initial condition $p(z_h=30~\rm{GeV}^{-1})=0$, that is, we set $p=0$ at $T\simeq0$. The energy density $\epsilon=-p+sT$ and the trace anomaly $\epsilon-3p$ are then obtained from the entropy and pressure density.

\begin{figure}[!h]
\begin{center}
\epsfxsize=7 cm \epsfysize=4.5 cm \epsfbox{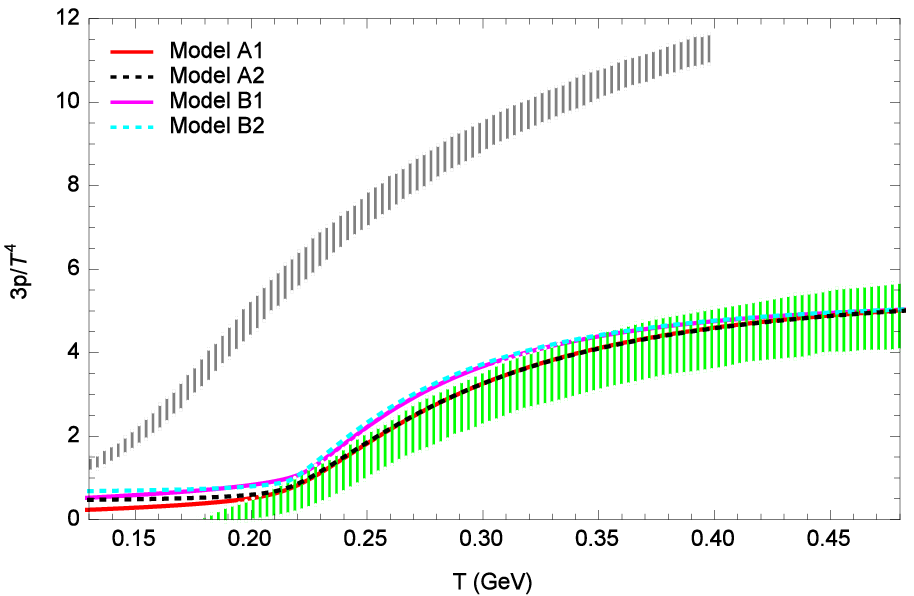}
\hspace*{0.6cm} \epsfxsize=7 cm \epsfysize=4.5 cm
\epsfbox{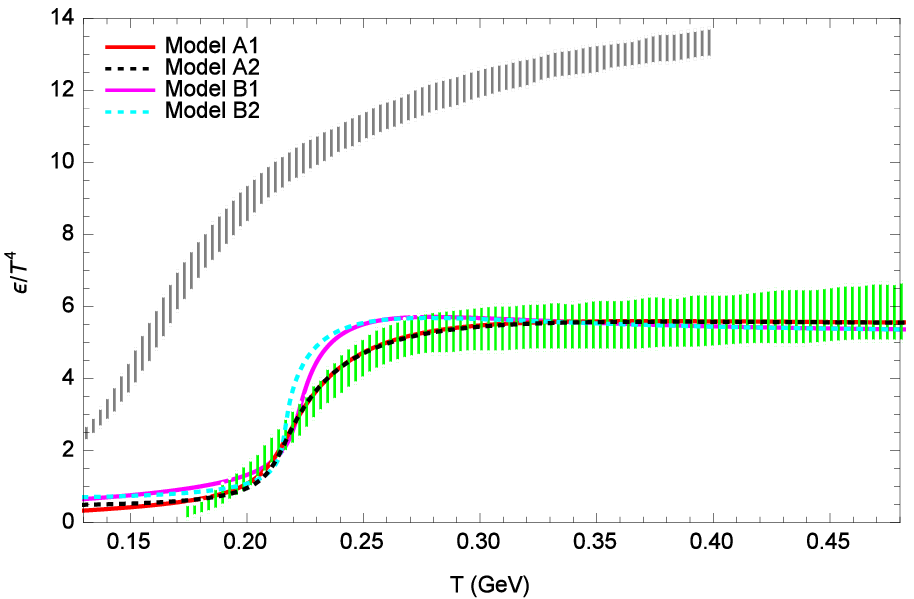} \vskip -0.05cm \hskip 0.7 cm
\textbf{(a)} \hskip 7.2 cm \textbf{(b)} \\
\end{center}
\caption{The scaled pressure density $3p/T^4$ and energy density $\epsilon/T^4$ as a function of $T$ in model A1 (A2) and B1 (B2). The green band is the lattice interpolation results with two flavors \cite{Burger:2014xga}, and the gray band is the lattice results with $2+1$ flavors \cite{Bazavov:2014pvz}.}
\label{p-e}
\end{figure}

\begin{figure}[!h]
\begin{center}
\epsfxsize=7 cm \epsfysize=4.5 cm \epsfbox{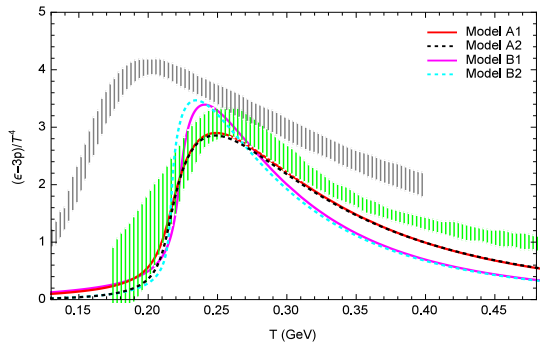}
\hspace*{0.8cm}  \vskip 0.1 cm
\end{center}
\caption{The scaled trace anomaly $(\epsilon-3p)/T^4$ as a function of $T$ in model A1 (A2) and B1 (B2). The green band is the lattice interpolation results with two flavors \cite{Burger:2014xga}, and the gray band is the lattice results with $2+1$ flavors \cite{Bazavov:2014pvz}.}
\label{anomaly}
\end{figure}

We show the numerical results in Fig.\ref{p-e} and Fig.\ref{anomaly}. The lattice results with two light flavors \cite{Burger:2014xga} are added in for fitting  and lattice results with $2+1$ flavors \cite{Bazavov:2014pvz} also added for comparison. In Fig.\ref{p-e}, we show the behavior of the scaled pressure and energy density with respect to temperature in the four models. The temperature dependence of the scaled trace anomaly is presented in Fig.\ref{anomaly}. The color band denotes the interpolation results of lattice simulations. One can see that the numerical results calculated from the four models are consistent with the $N_f=2$ lattice results and a similar crossover behavior emerged in a small region of temperature, which is consistent with the behavior of entropy density and sound speed in Fig.\ref{s-cs2}. The scaled pressure and energy density, which is associated with the degrees of freedom, increase with temperature gradually. At high temperature, the trace anomaly goes to zero, which indicates the system reaches asymptotically to the conformal gas. Qualitatively all the models give consistent results compared with the lattice data, and yet quantitatively the model A1 and A2 fit the lattice results much better than the model B1 and B2.

\section{Chiral and deconfining phase transition} \label{phasetransition}

In the previous section, we have already seen that the holographic model constructed in Sec.\ref{ED} can describe the equation of state in $N_f=2$ thermal QCD quite well. Qualitatively, these results signal a crossover transition in a similar temperature region. To test the hQCD model further and get more information of the thermal transition, we study the temperature dependent behavior of the order parameters for chiral and deconfining phase transitions in this section.

\subsection{Deconfining phase transition from Polyakov loop}
\label{deconfine}

The Polyakov loop operator is a useful quantity to investigate the phenomenon of deconfinement. It is defined as
\begin{equation}\label{Ploop}
L(T)=\frac{1}{N_c}\text{TrP exp}\left(ig\int_0^{1/T} \hat{A}_0(\tau,\vec{x})\,d\tau \right)
\end{equation}
with $N_c$ the color number, P denoting path ordering and the trace computed in the fundamental representation. As an order parameter for the center symmetry of the gauge group, its expectation value $\langle L(T)\rangle$ vanishes in the confined phase, as the center symmetry guarantees, while in the deconfined phase $\langle L(T)\rangle \neq 0$, which implies the center symmetry is breaking. In the holographic framework, the Polyakov loop is related to the action of the worldsheet which wraps the imaginary time circle. Schematically, we can write
\begin{equation}\label{VEV-Ploop}
\langle L(T)\rangle=\int DX\,e^{-S_w}.
\end{equation}

In the large $N_c$ and strong coupling limit, the expectation value of the Polyakov loop $\langle L(T)\rangle \sim e^{-S_{\rm{NG}}}$ with $S_{\rm{NG}}$ the Nambu-Goto action for the string worldsheet:
\begin{equation} \label{S-NG-Polyakov}
S_{\rm{NG}} =  \frac{1}{2\pi\alpha_p}\int d^2\eta \sqrt{  \rm{det} (g_{\mu\nu}^s  X_a^{\mu} X_b^{\nu})  } ,
\end{equation}
where $\alpha_p$ is the string tension, $g_{\mu\nu}^s$ the metric in the string frame, and $X_a^{\mu}$ the embedding function of the worldsheet in the target space. $\mu,\nu$ are $5D$ space-time indices and $a,b$ denote the worldsheet coordinates. Using the string frame metric (\ref{stringmetric}), we yield
\begin{equation}\label{S-NG2}
S_{\rm{NG}}=\frac{g_p }{\pi T}\int^{z_h}_0 dr\,\frac{
e^{2A_s}}{z^2} \sqrt{1+f(z) (\vec{x}\,')^2}
\end{equation}
with $g_p=\frac{L^2}{2\alpha_p}$. The prime indicates the derivative with respect to z. From the action $S_{\rm{NG}}$, the equation of motion for $\vec{x}$ is derived as
\begin{equation}\label{eq-x}
\biggl[\frac{ e^{2A_s}}{z^2}
f(z)\vec{x}\,'/\sqrt{1+f(z)(\vec{x}\,')^2}\biggr]'=0 \,.
\end{equation}

By substituting the constant solution of the above equation in the action $S_{\rm{NG}}$, the minimal worldsheet is obtained as
\begin{equation}\label{S-Ploop}
S_0 = c_p + S_0' = c_p + \frac{g_p}{\pi T} \int_0^{z_h} d z \left(\frac{e^{2A_s}}{z^2}- \frac{1}{z^2}\right),
\end{equation}
where $c_p$ is a normalization constant depending on the regularization scheme. Note that the last term of the integrand is a counter term which regularize the UV divergence of the original integral. Now we get the expectation value of the Polyakov loop
\begin{equation}\label{VEV-Ploop2}
\langle L(T) \rangle = w e^{-S_0} = e^{C_p - S_0' }
\end{equation}
with $C_p$ another normalization constant and $w$ a weight coefficient.

Using Eqs.(\ref{S-Ploop})-(\ref{VEV-Ploop2}) and the previous results fixed by the equation of state, we can obtain the temperature dependent behavior of Polyakov loop. The parameters $C_p, g_p$ are fixed by two-flavor lattice results from \cite{Burger:2011zc}. We present the numerical results of the expectation value of Polyakov loop as a function of $T$ in Fig.\ref{Ploop} (a). The corresponding parameters we used in the plot are listed in Table.\ref{ploopparameter}. One see that the expectation value of Polyakov loop increase from zero to finite value continuously with the increasing temperature, which shows a crossover transition from confined phase to deconfined phase. In addition, we find again that the model A1 and A2 fit the lattice data much better than the model B1 and B2, which is consistent with the study of equaiton of state.
\begin{table}[!h]
\begin{center}
\begin{tabular}{cccccccccc}
\hline\hline
                   &   $C_p$   &  $g_p$        \\
\hline
       Model A1   &   -0.3    &   0.9           \\

       Model A2     &   -0.25   &   0.86        \\

       Model B1   &   -1.2    &   1.6           \\

       Model B2     &    -1.2   &   1.6           \\
\hline\hline
\end{tabular}
\caption{Polyakov loop related parameters in model A1 (A2) and model B1 (B2).}
\label{ploopparameter}
\end{center}
\end{table}

Since in the case of finite quark mass there is not a real phase transition but a crossover one, the transition temperature $T_d$ is usually defined at the temperature where $\langle L \rangle$ changes fastest, i.e., the temperature with maximal $\frac{d\langle L \rangle}{dT}$. Therefore, we plot the derivative of $\langle L\rangle$ with respect to $T$ in Fig.\ref{Ploop} (b), from which the pseudo-critical temperature can be read from the location of the peak. We see that $T_d\simeq217~$MeV in model A1, A2 and B2, while in Model B1 $T_d\simeq223~$MeV. These results are consistent with the study of equation of state and comparable with the lattice results in \cite{Burger:2011zc}.
\begin{figure}[!h]
\begin{center}
\epsfxsize=7 cm \epsfysize=4.5 cm \epsfbox{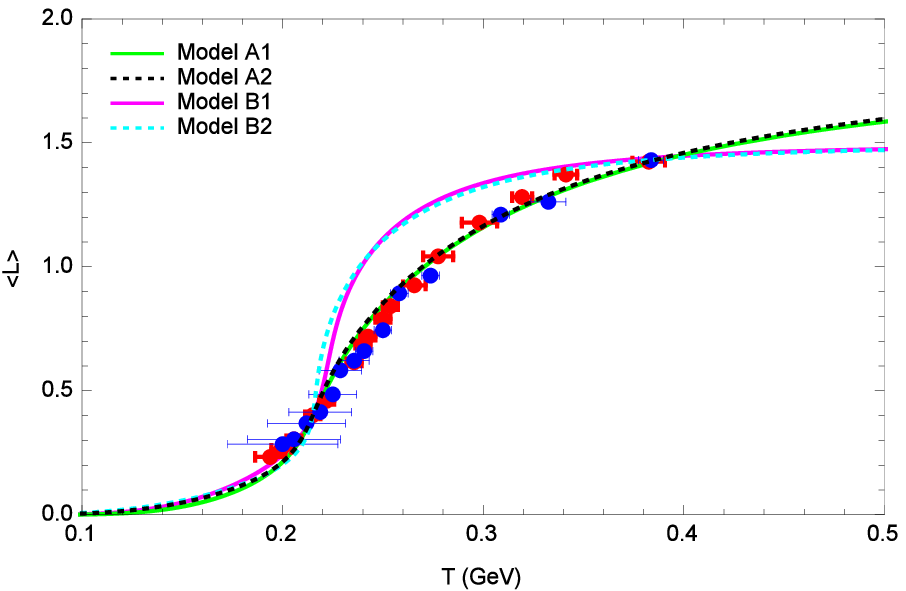}
\hspace*{0.6cm} \epsfxsize=7 cm \epsfysize=4.5 cm
\epsfbox{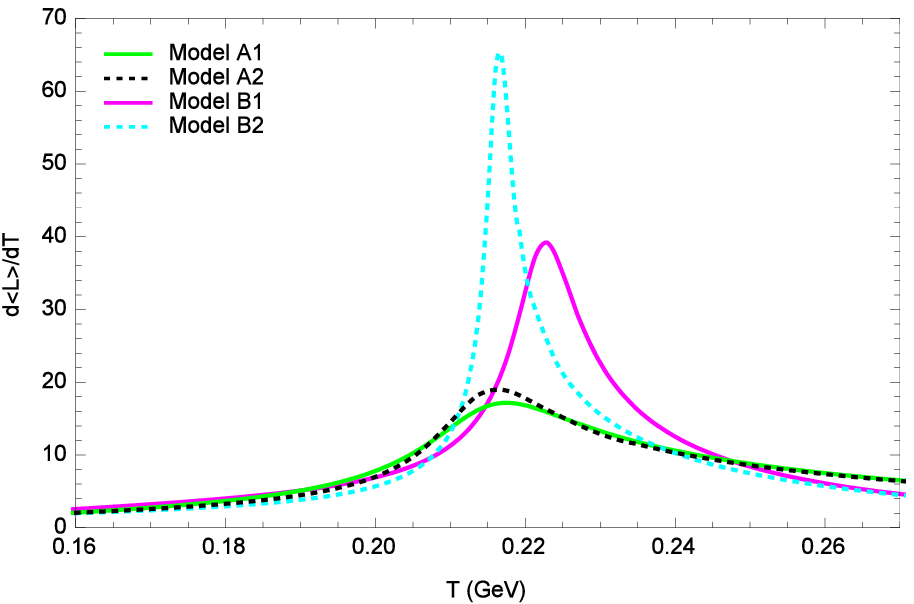} \vskip -0.05cm \hskip 0.7 cm
\textbf{(a)} \hskip 7.2 cm \textbf{(b)} \\
\end{center}
\caption{The expectation value of the Polyakov loop $\langle L(T)\rangle$ (a) and its derivative $\frac{d\langle L \rangle}{dT}$ with respect to $T$ (b) in model A1 (A2) and model B1 (B2). The red and blue points with error bars are two-flavor lattice data taken from \cite{Burger:2011zc}.  }
\label{Ploop}
\end{figure}

\subsection{Chiral phase transition from chiral condensate}\label{chiral part}

In the previous sections, we have constrained our models by studying the equation of state and have tested these ones further in the Polyakov loop calculation. In this section, we will go further and try to investigate another important aspect of QCD phase transition, i.e., the chiral phase transition.

As mentioned above, chiral symmetry breaking and restoration are characterized by the order parameter, i.e., chiral condensate $\sigma \equiv \langle \bar{q}q \rangle$, which can be encoded in the scalar field $X$ in the matter action (\ref{matter action}) \cite{Karch:2006pv}. The matter action $S_m$ has a $SU(2)_L\times SU(2)_R$ symmetry which is spontaneously broken when the scalar field $X$ obtains a vacuum expectation value $X_0$. Without loss of generality, we consider the case with degenerate quark mass, i.e., $m_u=m_d=m_q$, then we can set $X_0(z)=\chi(z) I_2/2$ with $I_2$ the $2\times2$ identity matrix. As the chiral condensate $\sigma$ only appears in $\chi(z)$, we derive the degenerate action of $\chi$ by inputting $X_0$ in the action (\ref{matter action}) as follows
\begin{eqnarray}\label{eff-action}
S_{\chi}=-\int d^5x \sqrt{-g^S}e^{-\phi}\left[\frac{1}{2}g^{zz}\chi^{'2}+V(\chi)\right],
\end{eqnarray}
where $V(\chi)\equiv \rm{Tr}(V_X(X))$. As noted in our previous work \cite{Chelabi:2015cwn,chiral-long}, the quartic term in the bulk scalar potential is necessary to realize the spontaneous chiral symmetry breaking. Hence, the potential $V(\chi)$ is given as
\begin{eqnarray}
V(\chi)=-\frac{3}{2}\chi^2+\lambda \chi^4.
\end{eqnarray}
Under the metric ansatz (\ref{stringmetric}), the equation of motion for $\chi$ is easily derived from Eq.(\ref{eff-action}) as
\begin{eqnarray}\label{eom-chi}
\chi^{''}(z)+\left(-\frac{3}{z}+3A_s^{'}(z)-\phi^{'}(z)+\frac{f^{'}(z)}{f(z)}\right)\chi^{'}(z)- \frac{e^{2A_s(z)}}{z^2f(z)}\partial_\chi V(\chi(z))=0,
\end{eqnarray}
from which the leading UV expansion of $\chi(z)$ can be obtained as follows
\begin{eqnarray}\label{chi-UV}
\chi(z)=m_q \zeta z+\frac{\sigma}{\zeta} z^3+\cdots,
\end{eqnarray}
where $\sigma$ is the chiral condensate and the normalization constant $\zeta=\frac{\sqrt{3}}{2\pi}$ \cite{Cherman:2008eh}.

At first sight, $m_q$ and $\sigma$ are two independent integral constant of Eq.(\ref{eom-chi}). Nevertheless, since $z=z_h$ is a singular point of Eq.(\ref{eom-chi}), the regular condition of $\chi$ would require $\frac{1}{f(z)}(f^{'}\chi^{'}-e^{2A_s}\partial_\chi V(\chi)/z^2)$ to be finite at $z=z_h$, which means $f^{'}\chi^{'}-e^{2A_s}\partial_\chi V(\chi)/z^2=0$ at horizon. This would impose an IR boundary condition naturally, and cause the chiral condensate to depend on the quark mass and temperature (For details, see \cite{Chelabi:2015cwn,chiral-long}). Using the above UV and IR boundary conditions, one can solve $\sigma$ as a function of $m_q$ and $T$.

\subsubsection{Spontaneous chiral symmetry breaking and its restoration}

\begin{figure}[!h]
\begin{center}
\epsfxsize=7 cm \epsfysize=4.5 cm \epsfbox{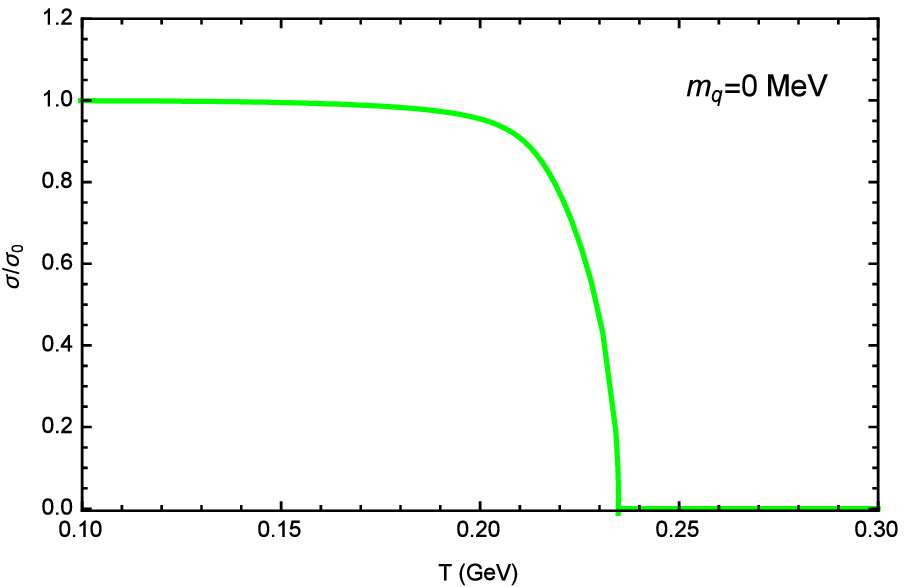}
\hspace*{0.6cm} \epsfxsize=7 cm \epsfysize=4.5 cm
\epsfbox{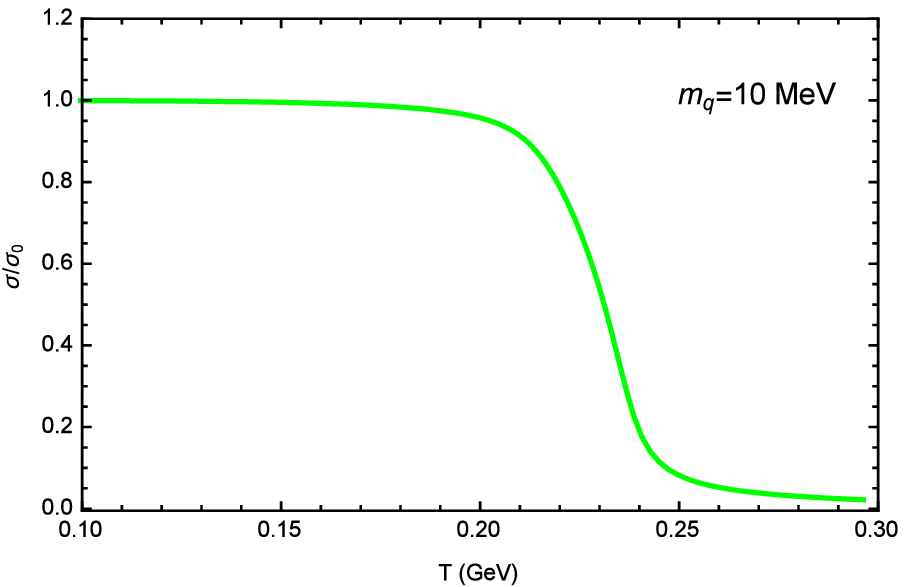} \vskip -0.05cm \hskip 0.7 cm
\textbf{(a)} \hskip 7.2 cm \textbf{(b)} \\
\end{center}
\caption{The temperature dependence of chiral condensate $\sigma$ in the chiral limit (a) and in finite quark mass case (b) in model A1.}
 \label{chiral phase}
\end{figure}

In the chiral limit $m_q=0$, there is always a trivial solution $\chi \equiv 0$ satisfying all the boundary conditions, which can be seen as the chiral symmetry restored phase. If there were chiral symmetry breaking in the system we studied, Eq.(\ref{eom-chi}) would have non-trivial solutions with $\chi \neq 0$ at some temperature region. Using the background and the parameters fixed in the previous sections, we do find non-trivial solutions below certain temperature for all the four models, which shows the validity of the constraints we used. Here we only take model A1 as an example and show the results in Fig.\ref{chiral phase} (a), in which we set $\lambda=8$.

In order to study the thermodynamic stability of the solutions, we need to compare free energies of the different solutions and select the lowest free energy branch. According to the AdS/CFT dictionary, the free energy can be approximated by the on-shell action shown as follows \cite{Chelabi:2015cwn,chiral-long}
\begin{eqnarray}\label{freeenergy}
\mathcal{F}\equiv \frac{F}{V_3} = -(\frac{1}{2z^3} \chi e^{3A_s-\phi}f\chi^{'})|_{\epsilon} - \lambda\int_0^{z_h} dz \sqrt{-g}e^{-\phi} \chi^4.
\end{eqnarray}
In the chiral limit, the first term in the above expression vanishes. It is apparent that if $\lambda>0$, the free energy of a non-trivial solution $\chi$ would be always smaller than the trivial solution $\chi=0$. That means the physical vacuum has non-zero chiral condensate, and the spontaneous chiral symmetry breaking is realized. Furthermore, from Fig.\ref{chiral phase} (a) one can see that the non-trivial solution disappears at around $235~\rm{MeV}$ with only the trivial solution left, which means that at high temperature the chiral symmetry is restored. It is easy to read from the figure that the phase transition is of second order, which is consistent with the sketch plot shown in Fig.\ref{columbia-plot}, since the derivative of $\sigma$ with respect to $T$ is discontinuous at the transition point.

Then we take finite $m_q$ to calculate $\sigma(T)$ and find that the qualitative results are all similar to the one shown in Fig.\ref{chiral phase} (b). However, when compared with Fig.\ref{chiral phase} (a), we find that the trivial solution $\chi=0$ has been disappeared as it does not satisfy the UV boundary condition (\ref{chi-UV}). The chiral condensate decreases very slowly at low temperature and goes through a sudden drop at certain temperature, then goes to zero slowly, which shows obviously a crossover transition. Thus, under our background models, we obtain second order chiral phase transition in the chiral limit and crossover transition in the case of finite quark mass. This is qualitatively coincident with the sketch in Fig.\ref{columbia-plot}\footnote{Here it should be noted that this is an approximate result, since in the full analysis combining $S_b$ and $S_m$, the quark mass would also affect the background.}.

\subsubsection{Confront chiral phase transition with lattice simulation}

The lattice QCD studies of chiral phase transition in recent years have attracted much attention, especially in the $2+1$ flavor case with physical quark mass \cite{Bazavov:2011nk,Aoki:2009sc,Borsanyi:2015waa,Bhattacharya:2014ara}. In \cite{Burger:2014xga,Burger:2011zc}, lattice simulation of thermal QCD transition with two light flavors was investigated for a set of pion masses ranging from $300~\rm{MeV}$ to $600~\rm{MeV}$. Therefore, it is also interesting to compare our model calculation with those lattice results. We roughly fit the chiral condensate related parameters $m_q,\lambda$ and present the results in Table.\ref{chiralparameters}.

\begin{table}[!h]
\begin{center}
\begin{tabular}{cccccccccc}
\hline\hline
                 &  $m_q (\rm{MeV})$  &  $\lambda$    \\
\hline
       Model A1    &       10      &        8      \\

       Model A2    &       20      &        8       \\

       Model B1   &        25      &        2        \\

       Model B2     &       25      &        2        \\
\hline\hline
\end{tabular}
\caption{Chiral condensate related parameters in model A1 (A2) and model B1 (B2).}
\label{chiralparameters}
\end{center}
\end{table}

\begin{figure}[!h]
\begin{center}
\epsfxsize=7 cm \epsfysize=4.5 cm \epsfbox{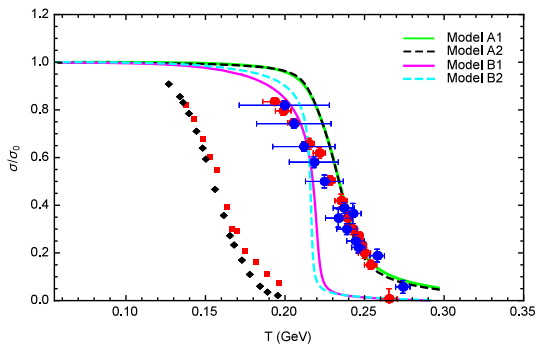}
\hspace*{0.8cm}  \vskip 0.1 cm
\end{center}
\caption{Comparison of chiral phase transition in model A1 (A2) and model B1 (B2) with lattice results. The red and blue filled circles are lattice data with two light flavors in different lattice extent \cite{Burger:2011zc}. The red and black filled squares are lattice data with $2+1$ flavors taken from \cite{Bazavov:2011nk}.}
\label{sigma compare}
\end{figure}

\begin{figure}[!h]
\begin{center}
\epsfxsize=7 cm \epsfysize=4.5 cm \epsfbox{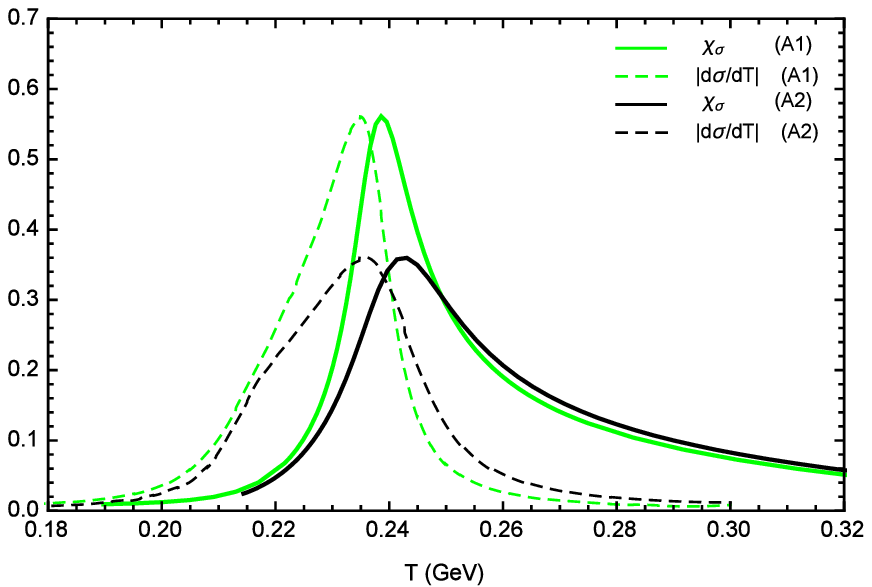}
\hspace*{0.6cm} \epsfxsize=7 cm \epsfysize=4.5 cm
\epsfbox{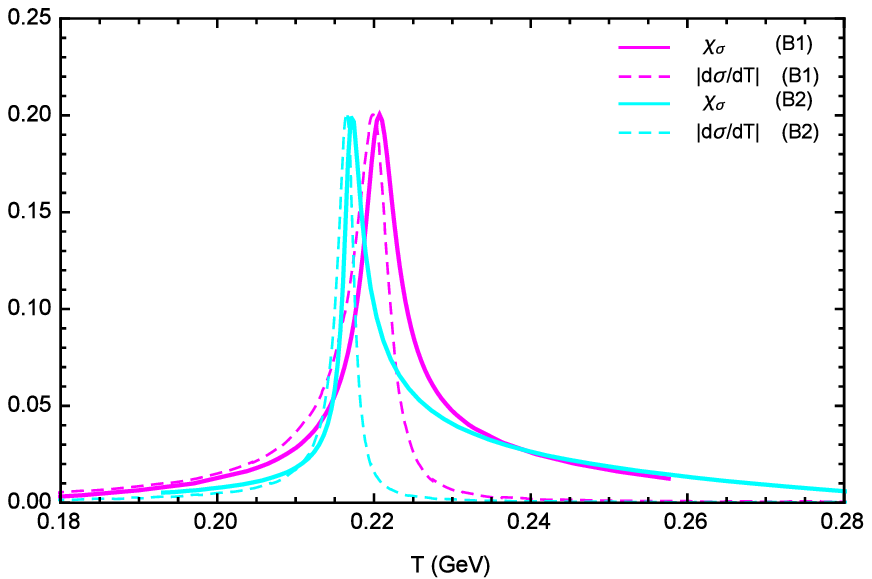} \vskip -0.05cm \hskip 0.7 cm
\textbf{(a)} \hskip 7.2 cm \textbf{(b)} \\
\end{center}
\caption{Chiral susceptibility $\chi_\sigma$ and the derivative of condensate with respect to temperature $\frac{d\sigma}{dT}$ in model A1 (A2) (a) and model B1 (B2) (b).}
\label{suscept-chiral}
\end{figure}

The temperature dependence of the chiral condensate in our models are shown in Fig.\ref{sigma compare}. One can see that the behavior of chiral condensates with $T$ in model A1 and A2 fit the lattice data better than that in model B1 and B2, which is consistent with the Polyakov loop and the equation of state calculations. The results of our models are much closer to the two-flavor lattice results than the ones of $2+1$ flavor, which is also consistent with our previous calculations. From Fig.\ref{sigma compare} one can also see that the chiral condensates decrease with temperature smoothly in the case of finite quark mass and go through a rapid drop when the temperature reaches to a value which characterizes the crossover transition of thermal QCD. To fix the chiral transition temperature $T_c$, we calculate the chiral susceptibility which is defined as
\begin{equation} \label{susceptibility}
\chi_\sigma=\frac{\partial\langle\bar{\psi}\psi\rangle}{\partial m_q}=\frac{\partial\sigma}{\partial m_q}.
\end{equation}

As a crosscheck, we also give another criterion to determine chiral transition temperature by the quantity $\frac{d\sigma}{dT}$. The pseudocritical temperature is identified as the extremal point of $\chi_\sigma$ or $\frac{d\sigma}{dT}$. The numerical results are presented in Fig.\ref{suscept-chiral}. One can see that although the curves of $\chi_\sigma$ and $\frac{d\sigma}{dT}$ are different overall, their extremal points lie in a narrow temperature region which indicates the possible transition temperature. The pseudocritical temperatures in model A1 and A2 are around $235\sim242~\rm{MeV}$, while the ones in model B1 and B2 are about $217\sim221~\rm{MeV}$. Compared with the previous results, one can see that the chiral transition temperature $T_c$ and the deconfining transition temperature $T_d$ in model B1 and B2 are much close with each other, while in model A1 and A2 $T_c$ seems a little higher than $T_d$. Fixing the two transition temperatures both from experimental side and from theoretical side is an interesting issue of many years and there are still many debates and discrepancies. We just show our results for a possible case.

\section{Discussion and conclusion}
\label{section-summary}

Thermal QCD studies of chiral and deconfining phase transitions have attracted much attention both from Lattice QCD \cite{Bhattacharya:2014ara,Burger:2014xga,Burger:2011zc,Bornyakov:2009qh,Bazavov:2011nk,Bazavov:2014pvz,Borsanyi:2013bia} and from effective field theories such as NJL models \cite{Ratti:2005jh,Roessner:2006xn,Rossner:2007ik}. In the holographic framework, the crossover behavior of the two phase transitions at physical quark mass has not yet been realized simultaneously. In this paper, we make the first attempt to accommodate these two phase transitions in a single holographic QCD model.

We consider the Einstein-dilaton system coupled with the soft-wall model action. As a preliminary try, we ignore the back-reaction from the flavor part to the background geometry. The flavor effects are taken into account by comparing the equation of state generating from the background part with those of two-flavor lattice simulation. By analyzing the Einstein-dilaton system, together with the constraints from the qualitative behavior of the two phase transitions, we propose four types of bottom-up holographic models (A1,A2,B1,B2). Then we use potential reconstruction approach to solve the whole background system. We find that within the physically concerned region, the temperature dependence of the dilaton potential is very weak, which imply the validity of the method we used. Under these backgrounds, we calculate the equation of state and compare with the two-flavor lattice QCD results. We find our results of pressure density, energy density and trace anomaly are in good agreement with the two-flavor lattice results in \cite{Burger:2014xga}. We also see the sudden release of degrees of freedom in the scaled entropy, energy and pressure density, which indicate the crossover behavior of the thermal transition.

Then we try to study temperature dependent behavior of Polyakov loop and chiral condensate, which are the corresponding order parameters of deconfining and chiral phase transition respectively. We find that all the four models give the correct qualitative behavior of Polyakov loop. All of them show a crossover transition from confined phase to deconfined phase with increasing temperature. Furthermore, the calculations in models A1 and A2 fit the two-flavor lattice results from \cite{Burger:2011zc} in very good accuracy. After that, we investigate chiral condensate in the matter part. We find that in the chiral limit, the spontaneous chiral symmetry breaking and its restoration are correctly realized. And then we take finite quark mass and compare the results with the two-flavor lattice ones from \cite{Burger:2011zc}. We find crossover behavior from a chiral symmetry breaking phase to a restored phase in the chiral transition in all the four models.

The transition temperature extracted from the location of the valley of sound speed square is around $217~\rm{MeV}$ for models A1, A2 and B2, while for model B1 is around $223~\rm{MeV}$. We note that this is very close to the one extracted from the peak of the derivative of Polyakov loop with respect to temperature, which is reasonable since the valley of sound speed square contains the information of degree of freedom in both two phases and the deconfining phase transition is related to the release of color degrees of freedom. This can be seen as one of the consistency checks in our models. We also extract the transition temperature from the chiral susceptibility $\chi_\sigma$ and the derivative of chiral condensate with respect to temperature. We note that the deviation of the transition temperatures extracted from the two ways is small, so a reasonable chiral transition temperature can be obtained. For the models A1 and A2 the transition temperatures are both around $235\sim242 ~\rm{MeV}$, which is slightly larger than the deconfining temperature, while for the models B1 and B2 they are around $217\sim221~\rm{MeV}$, which is very close to the corresponding deconfining temperature.

Finally, we emphasize that the studies here are very preliminary, and the main motivation here is to grasp the necessary ingredients in characterizing the two phase transitions simultaneously. We have not considered the back-reaction of the flavor part to the background geometry. We also note that the models A1 and A2 give better fittings to the lattice results in all the quantities, including the equation of state, the Polyakov loop and the chiral condensate. In this sense, the dominant effects in the intermediate region might be the $z^2$ term rather than $z^4$. These studies could provide clues for the more careful research in the future. It is also quite interesting to study the chemical potential effects in these models, which would be finished in the near future.

\vskip 1cm \noindent {\bf Acknowledgments}:
We are grateful to Ronggen Cai, Mei Huang, Shigeki Sugimoto, Tadashi Takayanagi, Yue-Liang Wu and Yi Yang for useful conversations and correspondence. Z.F. thanks supervisor Yue-Liang Wu for his patient instruction. S.H. thanks Tadashi Takayanagi for their encouragement and support. S.H. is supported by JSPS postdoctoral fellowship for foreign researchers and by the National Natural Science Foundation of China (No.11305235). This work is funded in part by China Postdoctoral Science Foundation.


\vspace*{5mm}




\begin{thebibliography}{99}

\bibitem{nature-PTD}
  Y.~Aoki, G.~Endrodi, Z.~Fodor, S.~D.~Katz and K.~K.~Szabo,
  Nature {\bf 443} (2006) 675
  [hep-lat/0611014].

\bibitem{qcd-phase-diagram}
  E.~Laermann and O.~Philipsen,
  Ann.\ Rev.\ Nucl.\ Part.\ Sci.\  {\bf 53} (2003) 163
  [hep-ph/0303042].

\bibitem{deForcrand:2006pv}
  P.~de Forcrand and O.~Philipsen,
  JHEP {\bf 0701} (2007) 077
  [hep-lat/0607017].

\bibitem{Kanaya:2010qd}
  K.~Kanaya,
  AIP Conf.\ Proc.\  {\bf 1343}, 57 (2011)
  doi:10.1063/1.3574942
  [arXiv:1012.4235 [hep-ph]].

\bibitem{Pisarski:1983ms}
  R.~D.~Pisarski and F.~Wilczek,
  Phys.\ Rev.\ D {\bf 29} (1984) 338.


\bibitem{Fischer:2009gk}
  C.~S.~Fischer and J.~A.~Mueller,
  Phys.\ Rev.\ D {\bf 80} (2009) 074029
  [arXiv:0908.0007 [hep-ph]].


\bibitem{Andersen:2013swa}
  J.~O.~Andersen, W.~R.~Naylor and A.~Tranberg,
  JHEP {\bf 1404}, 187 (2014)
  [arXiv:1311.2093 [hep-ph]].

\bibitem{Fischer:2011mz}
  C.~S.~Fischer, J.~Luecker and J.~A.~Mueller,
  Phys.\ Lett.\ B {\bf 702} (2011) 438
  [arXiv:1104.1564 [hep-ph]].

\bibitem{Xu:2011pz}
  F.~Xu, H.~Mao, T.~K.~Mukherjee and M.~Huang,
  Phys.\ Rev.\ D {\bf 84} (2011) 074009
  [arXiv:1104.0873 [hep-ph]].

\bibitem{Maldacena:1997re}
  J.~M.~Maldacena,
  Adv.\ Theor.\ Math.\ Phys.\  {\bf 2}, 231 (1998)
  [Int.\ J.\ Theor.\ Phys.\  {\bf 38}, 1113 (1999)]  [arXiv:hep-th/9711200].

\bibitem{Gubser:1998bc}
  S.~S.~Gubser, I.~R.~Klebanov and A.~M.~Polyakov,
  Phys.\ Lett.\  B {\bf 428}, 105 (1998)
  [arXiv:hep-th/9802109].

\bibitem{Witten:1998qj}
  E.~Witten,
  Adv.\ Theor.\ Math.\ Phys.\  {\bf 2}, 253 (1998)
  [arXiv:hep-th/9802150].

 \bibitem{Erlich:2005qh}
  J.~Erlich, E.~Katz, D.~T.~Son, M.~A.~Stephanov,
  Phys.\ Rev.\ Lett.\  {\bf 95}, 261602 (2005).
  [hep-ph/0501128].


 \bibitem{TB:05} G.~F.~de Teramond and S.~J.~Brodsky,
Phys.\ Rev.\ Lett.\ \textbf{94}, 201601 (2005).  [arXiv:hep-th/0501022].

\bibitem{DaRold2005} L.~Da Rold and A.~Pomarol,
Nucl.\ Phys.\ B \textbf{721}, 79 (2005).  [arXiv:hep-ph/0501218].

\bibitem{D3-D7} J.~Babington, J.~Erdmenger, N.~J.~Evans, Z.~Guralnik and I.~Kirsch,
Phys.\ Rev.\ D \textbf{69}, 066007 (2004)  [arXiv:hep-th/0306018].

\bibitem{D4-D6}
  M.~Kruczenski, D.~Mateos, R.~C.~Myers and D.~J.~Winters,
  JHEP {\bf 0405} (2004) 041
  doi:10.1088/1126-6708/2004/05/041
  [hep-th/0311270].

\bibitem{SS-1} T.~Sakai and S.~Sugimoto,
Prog.\ Theor.\ Phys.\ \textbf{113}, 843 (2005);  [arXiv:hep-th/0412141].

 \bibitem{SS-2}
 T.~Sakai and S.~Sugimoto,
'' Prog.\ Theor.\ Phys.\ \textbf{114}, 1083 (2006).  [arXiv:hep-th/0507073].

\bibitem{Csaki:2006ji}
  C.~Csaki and M.~Reece,
  ``Toward a systematic holographic QCD: A braneless approach,''
  JHEP {\bf 0705}, 062 (2007)
  [arXiv:hep-ph/0608266].

\bibitem{Dp-Dq}
 S.~He, M.~Huang, Q.~S.~Yan and Y.~Yang,
  Eur.Phys.J.C.(2010)66:187.
  arXiv:0710.0988 [hep-ph].

\bibitem{Gherghetta-Kapusta-Kelley}
T.~Gherghetta, J.~I.~Kapusta and T.~M.~Kelley,
Phys.\ Rev.\ D {\bf 79} (2009) 076003;  

\bibitem{Gherghetta-Kapusta-Kelley-2}
T.~M.~Kelley, S.~P.~Bartz and J.~I.~Kapusta,
  Phys.\ Rev.\ D {\bf 83} (2011) 016002;

\bibitem{YLWu}
  Y.~-Q.~Sui, Y.~-L.~Wu, Z.~-F.~Xie and Y.~-B.~Yang,
  Phys.\ Rev.\ D {\bf 81} (2010) 014024;  

\bibitem{YLWu-1}
 Y.~-Q.~Sui, Y.~-L.~Wu and Y.~-B.~Yang,
  Phys.\ Rev.\ D {\bf 83} (2011) 065030.  


\bibitem{Cui:2013xva}
  L.~X.~Cui, Z.~Fang and Y.~L.~Wu,
  arXiv:1310.6487 [hep-ph].

\bibitem{Li:2012ay}
  D.~Li, M.~Huang and Q.~S.~Yan,
  Eur.\ Phys.\ J.\ C {\bf 73} (2013) 2615
  [arXiv:1206.2824 [hep-th]].


\bibitem{Li:2013oda}
  D.~Li and M.~Huang,
  JHEP {\bf 1311} (2013) 088
  [arXiv:1303.6929 [hep-ph]].


\bibitem{Cherman:2008eh}
 A.~Cherman, T.~D.~Cohen and E.~S.~Werbos,
  Phys.\ Rev.\ C {\bf 79}, 045203 (2009)
  [arXiv:0804.1096 [hep-ph]].


\bibitem{Shuryak:2004cy}
  E.~V.~Shuryak,
  Nucl.\ Phys.\  A {\bf 750}, 64 (2005)
  [arXiv:hep-ph/0405066];

\bibitem{Tannenbaum:2006ch}
  M.~J.~Tannenbaum,
  Rept.\ Prog.\ Phys.\  {\bf 69}, 2005 (2006)
  [arXiv:nucl-ex/0603003].

\bibitem{Policastro:2001yc}
  G.~Policastro, D.~T.~Son and A.~O.~Starinets,
  Phys.\ Rev.\ Lett.\  {\bf 87}, 081601 (2001)
  [arXiv:hep-th/0104066].

\bibitem{Cai:2009zv}
  R.~-G.~Cai, Z.~-Y.~Nie, N.~Ohta and Y.~-W.~Sun,
 Phys.\ Rev.\ D {\bf 79}, 066004 (2009)  [arXiv:0901.1421 [hep-th]].  

\bibitem{Cai:2008ph}
  R.~-G.~Cai, Z.~-Y.~Nie and Y.~-W.~Sun,
   Phys.\ Rev.\ D {\bf 78}, 126007 (2008)  [arXiv:0811.1665 [hep-th]].  


\bibitem{Sin:2004yx}
  S.~J.~Sin and I.~Zahed,
  Phys.\ Lett.\  B {\bf 608}, 265 (2005)
  [arXiv:hep-th/0407215];\\

\bibitem{Shuryak:2005ia}
  E.~Shuryak, S.~J.~Sin and I.~Zahed,
  J.\ Korean Phys.\ Soc.\  {\bf 50}, 384 (2007)
  [arXiv:hep-th/0511199].

\bibitem{Nastase:2005rp}
  H.~Nastase,
  arXiv:hep-th/0501068.


\bibitem{Janik:2005zt}
  R.~A.~Janik and R.~B.~Peschanski,
  Phys.\ Rev.\  D {\bf 73}, 045013 (2006)
  [arXiv:hep-th/0512162];

\bibitem{Nakamura:2006ih}
  S.~Nakamura and S.~J.~Sin,
  JHEP {\bf 0609}, 020 (2006)
  [arXiv:hep-th/0607123];\\

\bibitem{Sin:2006pv}
  S.~J.~Sin, S.~Nakamura and S.~P.~Kim,
  JHEP {\bf 0612}, 075 (2006)
  [arXiv:hep-th/0610113].

\bibitem{Herzog:2006gh}
  C.~P.~Herzog, A.~Karch, P.~Kovtun, C.~Kozcaz and L.~G.~Yaffe,
  JHEP {\bf 0607}, 013 (2006)
  [arXiv:hep-th/0605158];

\bibitem{Gubser-drag}
  S.~S.~Gubser, 
  Phys.\ Rev.\  D {\bf 74}, 126005 (2006)
  [arXiv:hep-th/0605182].

\bibitem{Zhang:2012jd}
  Z.~q.~Zhang, D.~f.~Hou and H.~c.~Ren,
  JHEP {\bf 1301} (2013) 032
  [arXiv:1210.5187 [hep-th]].

\bibitem{Wu:2014gla}
  Y.~Wu, D.~Hou and H.~c.~Ren,
  arXiv:1401.3635 [hep-ph].

\bibitem{Li:2014hja}
  ``Enhancement of jet quenching around phase transition: result from the dynamical holographic model,''
  Phys.\ Rev.\ D {\bf 89} (2014) 12,  126006
  [arXiv:1401.2035 [hep-ph]].

\bibitem{Li:2014dsa}
  D.~Li, S.~He and M.~Huang,
  JHEP {\bf 1506} (2015) 046
  [arXiv:1411.5332 [hep-ph]].



\bibitem{Aharony:1999ti}
  O.~Aharony, S.~S.~Gubser, J.~M.~Maldacena, H.~Ooguri and Y.~Oz,
  Phys.\ Rept.\  {\bf 323}, 183 (2000)
  [arXiv:hep-th/9905111].


\bibitem{Erdmenger:2007cm}
  J.~Erdmenger, N.~Evans, I.~Kirsch and E.~Threlfall,
  Eur.\ Phys.\ J.\ A {\bf 35} (2008) 81
  [arXiv:0711.4467 [hep-th]].

\bibitem{Brodsky:2014yha}
  S.~J.~Brodsky, G.~F.~de Teramond, H.~G.~Dosch and J.~Erlich,
  Phys.\ Rept.\  {\bf 584} (2015) 1
  [arXiv:1407.8131 [hep-ph]].


\bibitem{Kim:2012ey}
  Y.~Kim, I.~J.~Shin and T.~Tsukioka,
  Prog.\ Part.\ Nucl.\ Phys.\  {\bf 68} (2013) 55
  [arXiv:1205.4852 [hep-ph]].

\bibitem{Adams:2012th}
  A.~Adams, L.~D.~Carr, T.~Schäfer, P.~Steinberg and J.~E.~Thomas,
  New J.\ Phys.\  {\bf 14} (2012) 115009
  [arXiv:1205.5180 [hep-th]].


\bibitem{Herzog:2006ra}
  C.~P.~Herzog,
  Phys.\ Rev.\ Lett.\  {\bf 98}, 091601 (2007)
  [hep-th/0608151].

\bibitem{BallonBayona:2007vp}
  C.~A.~Ballon Bayona, H.~Boschi-Filho, N.~R.~F.~Braga and L.~A.~Pando Zayas,
  Phys.\ Rev.\ D {\bf 77} (2008) 046002
  [arXiv:0705.1529 [hep-th]].

\bibitem{Cai:2007zw}
  R.~G.~Cai and J.~P.~Shock,
  JHEP {\bf 0708} (2007) 095
  [arXiv:0705.3388 [hep-th]].

\bibitem{Cai:2012eh}
  R.~G.~Cai, S.~Chakrabortty, S.~He and L.~Li,
  JHEP {\bf 1302} (2013) 068
  [arXiv:1209.4512 [hep-th]].

\bibitem{Kim:2007em}
  Y.~Kim, B.~H.~Lee, S.~Nam, C.~Park and S.~J.~Sin,
  Phys.\ Rev.\ D {\bf 76} (2007) 086003
  [arXiv:0706.2525 [hep-ph]].

\bibitem{Kim:2009wt}
  Y.~Kim, T.~Misumi and I.~J.~Shin,
  arXiv:0911.3205 [hep-ph].

\bibitem{Andreev-T3}
  O.~Andreev,
  Phys.\ Rev.\ Lett.\  {\bf 102}, 212001 (2009)
  [arXiv:0903.4375 [hep-ph]].

\bibitem{Colangelo:2010pe}
  P.~Colangelo, F.~Giannuzzi and S.~Nicotri,
  Phys.\ Rev.\ D {\bf 83} (2011) 035015
  [arXiv:1008.3116 [hep-ph]].



\bibitem{Gubser:2008yx}
  S.~S.~Gubser, A.~Nellore, S.~S.~Pufu and F.~D.~Rocha,
  Phys.\ Rev.\ Lett.\  {\bf 101} (2008) 131601
  [arXiv:0804.1950 [hep-th]].

\bibitem{Gubser:2008ny}
  S.~S.~Gubser and A.~Nellore,
  Phys.\ Rev.\ D {\bf 78} (2008) 086007
  [arXiv:0804.0434 [hep-th]].

\bibitem{Gubser:2008sz}
  S.~S.~Gubser, S.~S.~Pufu and F.~D.~Rocha,
  JHEP {\bf 0808}, 085 (2008).



\bibitem{Gursoy:2008bu}
  U.~Gursoy, E.~Kiritsis, L.~Mazzanti and F.~Nitti,
  Phys.\ Rev.\ Lett.\  {\bf 101} (2008) 181601
  [arXiv:0804.0899 [hep-th]].

\bibitem{Gursoy}
  U.~Gursoy and E.~Kiritsis,
  JHEP {\bf 0802}, 032 (2008);
  [arXiv:0707.1324 [hep-th]].

\bibitem{Gursoy-3}

  U.~Gursoy, E.~Kiritsis and F.~Nitti,
  JHEP {\bf 0802}, 019 (2008).
  [arXiv:0707.1349 [hep-th]].



\bibitem{Gursoy:2008za}
  U.~Gursoy, E.~Kiritsis, L.~Mazzanti and F.~Nitti,
  JHEP {\bf 0905} (2009) 033
  [arXiv:0812.0792 [hep-th]].


\bibitem{Finazzo:2014zga}
  S.~I.~Finazzo and J.~Noronha,
  Phys.\ Rev.\ D {\bf 90} (2014) 11,  115028
  [arXiv:1411.4330 [hep-th]].

\bibitem{Yaresko:2013tia}
  R.~Yaresko and B.~Kampfer,
  Phys.\ Lett.\ B {\bf 747} (2015) 36
  [arXiv:1306.0214 [hep-ph]].

\bibitem{Yaresko:2015ysa}
  R.~Yaresko, J.~Knaute and B.~Kämpfer,
  Eur.\ Phys.\ J.\ C {\bf 75} (2015) 6,  295
  [arXiv:1503.09065 [hep-ph]].



\bibitem{Li:2011hp}
  D.~Li, S.~He, M.~Huang and Q.~S.~Yan,
  JHEP {\bf 1109} (2011) 041
  [arXiv:1103.5389 [hep-th]].

\bibitem{Cai:2012xh}
  R.~G.~Cai, S.~He and D.~Li,
  JHEP {\bf 1203} (2012) 033
  [arXiv:1201.0820 [hep-th]].



\bibitem{He:2013qq}
  S.~He, S.~Y.~Wu, Y.~Yang and P.~H.~Yuan,
  JHEP {\bf 1304} (2013) 093
  [arXiv:1301.0385 [hep-th]].

\bibitem{Yang:2014bqa}
  Y.~Yang and P.~H.~Yuan,
  JHEP {\bf 1411}, 149 (2014)
  [arXiv:1406.1865 [hep-th]].

\bibitem{Yang:2015aia}
  Y.~Yang and P.~H.~Yuan,
  arXiv:1506.05930 [hep-th].


\bibitem{Zuo:2014iza}
  F.~Zuo,
  JHEP {\bf 1406} (2014) 143
  [arXiv:1404.4512 [hep-ph]].

\bibitem{Zuo:2014vga}
  F.~Zuo and Y.~H.~Gao,
  JHEP {\bf 1407} (2014) 147
  [arXiv:1403.2241 [hep-ph]].

\bibitem{Cui:2014oba}
  L.~X.~Cui, Z.~Fang and Y.~L.~Wu,
  arXiv:1404.0761 [hep-ph].



\bibitem{Afonin:2014jha}
  S.~S.~Afonin and A.~D.~Katanaeva,
  Eur.\ Phys.\ J.\ C {\bf 74}, no. 10, 3124 (2014)
  [arXiv:1408.6935 [hep-ph]].



\bibitem{Karch:2006pv}
  A.~Karch, E.~Katz, D.~T.~Son and M.~A.~Stephanov,
  Phys.\ Rev.\ D {\bf 74}, 015005 (2006)
  [hep-ph/0602229].

\bibitem{Nambu:1961tp}
  Y.~Nambu and G.~Jona-Lasinio,
  Phys.\ Rev.\  {\bf 122}, 345 (1961).



\bibitem{Nambu:1961fr}
  Y.~Nambu and G.~Jona-Lasinio,
  Phys.\ Rev.\  {\bf 124} (1961) 246.

\bibitem{Colangelo:2008us}
  P.~Colangelo, F.~Giannuzzi, S.~Nicotri and V.~Tangorra,
  Eur.\ Phys.\ J.\ C {\bf 72} (2012) 2096
  doi:10.1140/epjc/s10052-012-2096-9
  [arXiv:1112.4402 [hep-ph]].


\bibitem{Chelabi:2015cwn}
  K.~Chelabi, Z.~Fang, M.~Huang, D.~Li and Y.~L.~Wu,
  arXiv:1511.02721 [hep-ph].

\bibitem{chiral-long}
  K.~Chelabi, Z.~Fang, M.~Huang, D.~Li and Y.~L.~Wu,
  in preparation.


\bibitem{He:2011hw}
  S.~He, Y.~-P.~Hu and J.~-H.~Zhang,
  JHEP {\bf 1112}, 078 (2011)
  [arXiv:1111.1374 [hep-th]].

\bibitem{He:2010ye}
  S.~He, M.~Huang, Q.~-S.~Yan,
  Phys.\ Rev.\  {\bf D83}, 045034 (2011).
  [arXiv:1004.1880 [hep-ph]].

\bibitem{Kajantie:2011nx}
  K.~Kajantie, M.~Krssak, M.~Vepsalainen and A.~Vuorinen,
  Phys.\ Rev.\ D {\bf 84} (2011) 086004
  [arXiv:1104.5352 [hep-ph]].


\bibitem{Batell:2008zm}
  B.~Batell and T.~Gherghetta,
  Phys.\ Rev.\ D {\bf 78}, 026002 (2008)
  [arXiv:0801.4383 [hep-ph]].


\bibitem{Gubarev:2000eu}
  F.~V.~Gubarev, L.~Stodolsky, V.~I.~Zakharov,
  Phys.\ Rev.\ Lett.\  {\bf 86}, 2220-2222 (2001).
  [hep-ph/0010057].

  \bibitem{Gubarev:2000nz}
  F.~V.~Gubarev, V.~I.~Zakharov,
  Phys.\ Lett.\  {\bf B501}, 28-36 (2001).
  [hep-ph/0010096].

  \bibitem{Kondo:2001nq}
  K.~I.~Kondo,
  Phys.\ Lett.\  B {\bf 514}, 335 (2001)
  [arXiv:hep-th/0105299].


\bibitem{Borsanyi:2015waa}
  S.~Borsanyi {\it et al.},
  Phys.\ Rev.\ D {\bf 92}, no. 1, 014505 (2015)
  doi:10.1103/PhysRevD.92.014505
  [arXiv:1504.03676 [hep-lat]].

\bibitem{Aoki:2009sc}
  Y.~Aoki, S.~Borsanyi, S.~Durr, Z.~Fodor, S.~D.~Katz, S.~Krieg and K.~K.~Szabo,
  JHEP {\bf 0906}, 088 (2009)
  doi:10.1088/1126-6708/2009/06/088
  [arXiv:0903.4155 [hep-lat]].

\bibitem{Bazavov:2011nk}
  A.~Bazavov {\it et al.},
  Phys.\ Rev.\ D {\bf 85} (2012) 054503
  [arXiv:1111.1710 [hep-lat]].

\bibitem{Bhattacharya:2014ara}
  T.~Bhattacharya {\it et al.},
  Phys.\ Rev.\ Lett.\  {\bf 113}, no. 8, 082001 (2014)
  doi:10.1103/PhysRevLett.113.082001
  [arXiv:1402.5175 [hep-lat]].

\bibitem{Burger:2014xga}
  F.~Burger {\it et al.} [tmfT Collaboration],
  Phys.\ Rev.\ D {\bf 91}, no. 7, 074504 (2015)
  [arXiv:1412.6748 [hep-lat]].

\bibitem{Burger:2011zc}
  F.~Burger {\it et al.} [tmfT Collaboration],
  Phys.\ Rev.\ D {\bf 87}, no. 7, 074508 (2013)
  [arXiv:1102.4530 [hep-lat]].


\bibitem{Bazavov:2014pvz}
  A.~Bazavov {\it et al.} [HotQCD Collaboration],
  Phys.\ Rev.\ D {\bf 90}, no. 9, 094503 (2014)
  [arXiv:1407.6387 [hep-lat]].


\bibitem{Borsanyi:2013bia}
  S.~Borsanyi, Z.~Fodor, C.~Hoelbling, S.~D.~Katz, S.~Krieg and K.~K.~Szabo,
  Phys.\ Lett.\ B {\bf 730}, 99 (2014)
  [arXiv:1309.5258 [hep-lat]].


\bibitem{Bornyakov:2009qh}
  V.~G.~Bornyakov, R.~Horsley, S.~M.~Morozov, Y.~Nakamura, M.~I.~Polikarpov, P.~E.~L.~Rakow, G.~Schierholz and T.~Suzuki,
  Phys.\ Rev.\ D {\bf 82}, 014504 (2010)
  [arXiv:0910.2392 [hep-lat]].


\bibitem{Ratti:2005jh}
  C.~Ratti, M.~A.~Thaler and W.~Weise,
  Phys.\ Rev.\ D {\bf 73}, 014019 (2006)
  [hep-ph/0506234].

\bibitem{Roessner:2006xn}
  S.~Roessner, C.~Ratti and W.~Weise,
  Phys.\ Rev.\ D {\bf 75}, 034007 (2007)
  [hep-ph/0609281].

\bibitem{Rossner:2007ik}
  S.~Roessner, T.~Hell, C.~Ratti and W.~Weise,
  Nucl.\ Phys.\ A {\bf 814}, 118 (2008)
  [arXiv:0712.3152 [hep-ph]].



\end{thebibliography}
\end{document}